\documentclass[utf8]{frontiersinFPHY_FAMS} 
\setcitestyle{square} 
\usepackage{url,hyperref,lineno,microtype,subcaption}
\usepackage[onehalfspacing]{setspace}
\def\keyFont{\fontsize{8}{11}\helveticabold}
\def\firstAuthorLast{Superconductivity of Nb by 
                     M. Zarea, H. Ueki, J.A. Sauls}
\def\Authors{Mehdi Zarea\,$^{1}$, Hikaru Ueki\,$^{1}$ and J.~A. Sauls\,$^{1,*}$}
\def\Address{$^{1}$Hearne Institute of Theoretical Physics, 
                   Department of Physics \& Astronomy, 
                   Louisiana State University, 
                   Baton Rouge , LA 70803, USA}  
\def\corrAuthor{J.A. Sauls}
\def\corrEmail{sauls@lsu.edu}
\newcommand{\kb}{k_{\text{B}}}

\newcommand{\be}{\begin{equation}}
\newcommand{\ee}{\end{equation}}
\newcommand{\ber}{\begin{eqnarray}}
\newcommand{\eer}{\end{eqnarray}}
\newcommand{\sgn}{\mbox{sgn}}
\newcommand{\eps}{\varepsilon}
\newcommand{\grad}{\mbox{\boldmath$\nabla$}}

\def\ket#1{\mbox{$\displaystyle\vert\,#1\,\rangle$}}
\def\bra#1{\mbox{$\displaystyle\langle\,#1\,\vert$}}

\def\cA{{\mathcal A}}
\def\cO{{\mathcal O}}
\def\cD{{\mathcal D}}

\def\cN{{\mathcal N}}
\def\cY{{\mathcal Y}}

\def\ve{{\bf e}}
\def\vp{{\bf p}}
\def\vq{{\bf q}}
\def\vv{{\bf v}}

\def\vr{{\bf r}}

\def\vu{{\bf u}}

\def\vR{{\bf R}}

\def\nicefrac#1#2{\genfrac{}{}{}{1}{#1}{#2}}

\def\ns{\negthickspace}

\def\whT{\widehat{T}}

\def\tone{\widehat{1}}

\newcommand{\whtaux}{{\widehat{\tau}_1}}

\newcommand{\whtauz}{{\widehat{\tau}_3}}

\def\whDelta{\widehat{\Delta}}

\def\Tr#1{\mbox{Tr}\big\{#1\big\}}

\newcommand{\mfF}{\mathfrak{F}}

\newcommand{\mfG}{\mathfrak{G}}

\newcommand{\whmfF}{\widehat{\mathfrak{F}}}
\newcommand{\whmfG}{\widehat{\mathfrak{G}}}

\def\whSigma{\widehat{\Sigma}}
\def\whSig{\widehat{\mbox{$\Sigma$}}}
\newcommand{\Del}{{\Delta}}
\newcommand{\uDel}{{\underline\Delta}}
\def\whDel{\widehat{\mbox{$\Delta$}}}
\def\sml{{\bf\textsf{s}}}
\def\commutator#1#2{\mbox{$\displaystyle\,\left[\,#1\,,\,#2\,\right]$}}
\begin{document}
\onecolumn
\firstpage{1}
\title{Effects of anisotropy and disorder on the superconducting properties of Niobium} 
\author[\firstAuthorLast ]{\Authors}
\address{}
\correspondance{}
\extraAuth{}
\maketitle
%------------------------------------------------------------------------------------
\begin{abstract}
We report results for the superconducting transition temperature and anisotropic energy gap for pure Niobium based on Eliashberg's equations and electron and phonon band structures computed from density functional theory.
The electronic band structure is used to construct the Fermi surface and calculate the Fermi velocity at each point on the Fermi surface.
The phonon bands are in excellent agreement with inelastic neutron scattering data. The corresponding phonon density of states and electron-phonon coupling define the electron-phonon spectral function, $\alpha^2F(\vp,\vp';\omega)$, and the corresponding electron-phonon pairing interaction, which is the basis for computing the superconducting properties.
The electron-phonon spectral function is good agreement with existing tunneling spectroscopy data except for the spectral weight of the longitudinal phonon peak at $\hbar\omega_{\text{LO}}=23\,\mbox{meV}$.
We obtain an electron-phonon coupling constant of $\lambda=1.057$, renormalized Coulomb interaction, $\mu^{\star}=0.218$ and transition temperature $T_c=9.33\,\mbox{K}$. The corresponding strong-coupling gap at $T=0$ is modestly enhanced, $\Delta_0=1.55\,\mbox{meV}$, compared to the weak-coupling BCS value $\Delta_0^{\text{wc}}=1.78\,\kb\,T_c= 1.43\,\mbox{meV}$.
The superconducting gap function exhibits substantial anisotropy on the Fermi surface.
We analyze the distribution of gap anisotropy and compute the suppression of the superconducting transition temperature using a self-consistent T-matrix theory for quasiparticle-impurity scattering to describe Niobium doped with non-magnetic impurities. 
We compare these results with experimental results on Niobium SRF cavities doped with Nitrogen impurities.
\tiny
\keyFont{\section{Keywords:} electronic structure, phonon structure, first principle DFT calculations, Eliashberg theory, electron-phonon mediated superconductivity, anisotropic superconductors, impurity scattering, pair-breaking}
\end{abstract}
%------------------------------------------------------------------------------------

\section{Introduction}

The electronic properties of Niobium (Nb) and its alloys are central to the development of superconducting radio frequency (SRF) cavity technology for particle accelerators, as well as applications to device technologies for quantum computing and sensing applications~\citep{gur12,rea16}.
In particular role of disorder in the low-power quantum limit for the performance of superconducting Nb SRF cavities is an active area of research~\citep{rom17,lee22}.
Nitrogen-doped Nb, with quality factors of order $Q\approx 10^{11}$ and accelerating gradients as high as $45\,\mbox{MV/m}$, is also the superconductor of choice for SRF cavities used for high-energy accelerators~\citep{gra17}.
However, even for state-of-the-art cavities there is room for improved performance, both in terms of the quality factor as well as the maximum accelerating gradient.
Impurities and structural defects, nano-scale inclusions and two-level tunneling centers all impact the electromagnetic response of the current carrying region near the vacuum-superconducting interface, sometimes in counter-intuitive ways~\citep{nga19}.
In order to obtain a deeper understanding of the multiple roles in which impurities and defects impact the performance of Nb superconducting cavities, films and devices we develop the theory of moderately disordered superconducting Nb starting from first-principles theory of pure Nb informed by experimental data for the metallic and superconducting properties of high-purity bulk Nb. 
This report focusses on the zero-field equilibrium superconducting properties of pure Nb obtained from Eliashberg's theory for electron-phonon mediated superconductivity~\citep{eli61} with electronic structure, phonon structure and the electron-phonon coupling obtained from density functional theory (DFT)~\citep{koh65,deg95}.
We then investigate the effects of impurity disorder on N-doped Nb.

The anisotropy of the electron-phonon coupling and angle-resolved density of states, and thus the pairing amplitude for different momenta on the Fermi surface, combined with quasiparticle-impurity scattering leads to violation of Anderson's theorem~\citep{and59}, and thus a suppression of the superconducting transition temperature that increases with the impurity scattering rate. 
The suppression of $T_c$ by non-magnetic impurity disorder on conventional anisotropy superconductors such as Nb provides an excellent diagnostic of the impurity scattering rate in films and cavities. 
In a separate report we build these results into a theory for the microwave response of Nb with impurity and surface disorder~\citep{uek22}.

%------------------------------------------------------------------------------

\section{Eilenberger/Eliahsberg Theory}\label{sec-Quasiclassical_Theory}

The results reported here are based on the strong-coupling theory of electron-phonon mediated superconductivity in metallic alloys as formulated by Eliashberg, Eilenberger and Larkin and Ovchinnikov~\citep{eli60,eli61,eil68,lar69}. We use the notation of Ref.~\citep{rai94} which includes development of the theory from the formal quantum field theoretical equations for interacting Fermi systems.
For equilibrium states of superconductors a central object of the theory is the $4\times 4$ Nambu matrix propagator,
\begin{equation}\label{eq-quasiclassical_G} 
\whmfG(\vp,\vr;\varepsilon_n) 
= 
\left(
\begin{array}{cc}
\hat\mfG(\vp,\vr;\varepsilon_n) 
& 
\hat\mfF(\vp,\vr;\varepsilon_n) 
\\
\underline{\hat\mfF}(\vp,\vr;\varepsilon_n) 
& 
\underline{\hat\mfG}(\vp,\vr;\varepsilon_n) 
\end{array}
\right)
\,.
\end{equation}
The diagonal element is the $2\times 2$ quasiparticle propagator, $\mfG_{\alpha\beta}(\vp,\vr;\varepsilon_n)$, for momenta $\vp$ on the Fermi surface and Matsubara energy, $\varepsilon_n=(2n+1)\pi T$. In general the propagator is a function of spatial position $\vr$, with matrix elements labeled by $\alpha\beta$ in $2\times 2$ spin space. The equal-time propagator defines the one-particle density matrix, from which all one-body observables can be calculated. Analytic continuation of the diagonal Matsubara propagator to the real energy axis also determines the retarded propagator, $\hat\mfG^{\text{R}}(\vp,\vr;\varepsilon)=\hat\mfG(\vp,\vr;i\varepsilon_n\rightarrow\varepsilon+i0^{+})$, from which the spin-averaged, local density of states for Bogoliubov quasiparticles with momenta and excitation energies near the Fermi surface can be computed, $\cN(\vp,\vr;\varepsilon)=-\frac{N_f}{2\pi}\Im\,\Tr{\hat\mfG^{\text{R}}(\vp,\vr;\varepsilon)}$, where $N_f$ is the normal-state density of states at the Fermi energy.

The off-diagonal element, $\mfF_{\alpha\beta}(\vp,\vr;\varepsilon_n)$, is the anomalous (Gorkov) pair propagator in the quasiclassical limit, $\hbar v_f/2\pi T_c \gg \lambda_f$, i.e. Cooper pair size large compared to the Fermi wavelength. The equal-time propagator defines the local Cooper pair amplitude. The lower components of the Nambu matrix define the propagators for hole-like quasiparticles and the conjugate anomalous propagator, both of which are related to $\hat\mfG$ and $\hat\mfF$ by Fermion exchange symmetry and particle-hole symmetry. See Ref.~\citep{rai94} for the standard definitions for the propagators and their symmetries.

\subsection{Eilenberger's Equations}

The quasiparticle and anomalous pair propagators, organized into $4\times 4$ Nambu matrices, obey Gorkov's equations~\citep{gor58}. Eilenberger transformed Gorkov's equations into a matrix transport-type equation for the matrix propagator~\citep{eil68}, 
\be\label{eq-Eilenberger_Transport}
\hspace*{-3mm}
\commutator{i\varepsilon_n\whtauz-\whSigma(\vp,\vr;\varepsilon_n)}
           {\whmfG(\vp,\vr;\varepsilon_n)}
+ 
i\hbar\vv_{\vp}\cdot\grad_{\vr}\whmfG 
= 0
\,.
\ee
In contrast to Gorkov's equation, which is a second-order differential equation with a unit source term originating from the Fermion anti-commutation relations, Eilenberger's equation is a homogeneous, first-order differential equation describing the evolution of the matrix propagator along classical trajectories in phase space ($\vp,\vr$) defined by the Fermi velocity, $\vv_{\vp}=\grad_{\vp}\varepsilon_{\vp}$, where $\varepsilon_{\vp}$ is excitation energy of a normal-state electronic quasiparticle relative to the Fermi energy.
Eilenberger's transport equation determines the \emph{equilibrium} propagator, including inhomogeneous states generated by an external magnetic field and/or a spatially varying pairing self-energy, $\whDelta(\vp,\vr)$.
The transport equation is supplemented by Eilenberger's normalization condition~\citep{eil68},
\be\label{eq-Eilenberger_Normalization}
\whmfG(\vp,\vr;\varepsilon_n)^2 = -\pi^2\,\tone
\,,
\ee
which enforces the spectral weight implied by the source term in Gorkov's equation.
The physical properties of a particular superconducting material are encoded in the self energy functional, $\whSigma(\vp,\vr;\varepsilon_n)$, that enters Eq.~\eqref{eq-Eilenberger_Transport}. The self energy includes corrections to the effective mass of electronic quasiparticles from the coupling to phonons, mean-field polarization corrections to external perturbations as well as the off-diagonal pairing self energy resulting from the Cooper instability.
The self energies can be classified by the expansion parameters of Fermi liquid theory, $\sml=\{k_{\text{B}} T/E_f, \hbar/p_f\xi_0, \hbar/\tau E_f, \ldots\}$, and are defined by a diagrammatic expansion in the Nambu matrix propagator for quasiparticles and Cooper pairs, $\whmfG(\vp;\varepsilon_n)$ and the phonon propagator, $\cD_{\nu}(\vq,\omega_m)$~\citep{rai86,rai94}. 

\begin{figure}[t]
\centering\includegraphics[width=0.6\textwidth]{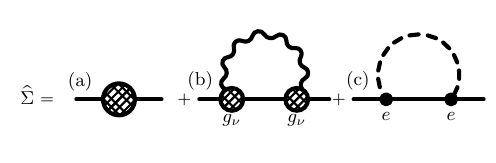}
\caption{
Leading order electronic self energy diagram
(a) determines the Fermi surface, Fermi velocity and electronic 
    contribution to the effective mass.
Next to leading order: diagram (b) is this exchange of a branch $\nu$ phonon of momentum $\vq$ and energy $\omega_{\nu}(\vq)$ represented by the phonon propagator, ${\mathcal D}_{\nu}(\vq,\omega_m)$, and the electron-phonon coupling $g_{\nu}$.
Diagram (c) is the electronic self energy generated by the screened Coulomb interaction, $V_{ee}(\vq,\omega_m)$ (dashed line). Intermediate states of quasiparticles and Cooper pairs are represented by the Nambu propagator, $\whmfG(\vp',\varepsilon_n')$ (solid line).}
\label{fig-Sigma-Eliashberg}
\end{figure}

The leading order contributions to the electronic self energy for elemental superconductors like Nb are shown in Fig.~\ref{fig-Sigma-Eliashberg}.
Diagram Fig.~\ref{fig-Sigma-Eliashberg}(a) is the leading order self energy, $\cO(\sml^0)$, that defines the Fermi level, Fermi surface and Fermi velocity in terms of bare electrons and their interactions. In Sec.~\ref{sec-DFT} we use the DFT code developed by Quantum Espresso~\citep{gia09,gia17} (QE) to determine the electronic bands in the low-energy region near the Fermi surface, the phonon band structure and the electron-phonon coupling strength for Niobium. 

%------------------- Leading Order Self Energies -------------------------
%------------------------------------------------------------------------------

\subsubsection{Electron-Phonon Self-Energy}

The electron-phonon self-energy in Fig.~\ref{fig-Sigma-Eliashberg}(b) provides the attractive interaction leading to superconductivity in most elemental superconductors~\citep{mig59,eli61}. Phonon-mediated interactions are retarded in time, and as a consequence the self energy, $\whSig_{\text{ep}}$, depends on energy in the low-energy phonon bandwidth. This leads to quasiparticle mass renormalization, finite quasiparticle lifetimes and electron-phonon mediated pairing interactions. The self energy corresponding to Fig.~\ref{fig-Sigma-Eliashberg}(b) is 
\be\label{eq-Self-Energy-Eliashberg}
\hspace*{-2mm}
\whSig_{\text{ep}}(\vp;\varepsilon_n)\ns 
=
\ns \int d\vp'\,T\sum_{n'}
\lambda(\vp,\vp';\varepsilon_{n}-\varepsilon_{n'})\,\whmfG(\vp';\varepsilon_{n'})
\,.
\ee
We use the short-hand notation $\vp$ to denote a point on the Fermi surface, which in general will have multiple sheets within the First Brillouin Zone (FBZ), $\int d\vp (\ldots) \equiv \int dS_{\vp}\,n(\vp)\,(\ldots)$ where the integral is over the area of the Fermi surface and $n(\vp)$ is the anisotropy of the normal-state density of states normalized to $\int d\vp (1) \equiv 1$.
For conventional spin-singlet pairing, and neglecting spin-dependent interactions, the electron-phonon self energy reduces to
\be
\hspace*{-2mm}
\whSig_{\text{ep}}(\vp;\varepsilon_n)\ns 
=
\begin{pmatrix}
\Sigma_{\text{ep}}(\vp;\varepsilon_n)\,\hat{1} 
& 
\Delta_{\text{ep}}(\vp;\varepsilon_n)\,i\sigma_y
\cr
\Delta^*_{\text{ep}}(\vp;-\varepsilon_n)\,i\sigma_y 
& 
-\Sigma_{\text{ep}}(\vp;\varepsilon_n)\,\hat{1}
\end{pmatrix}
\,.
\ee
The diagonal and off-diagonal self energies are given by 
\ber\label{eq-Self-Energy-ep-diagonal}
\hspace*{-2mm}
\Sigma_{\text{ep}}(\vp;\varepsilon_n)\ns 
=
\ns \int d\vp'\,T\sum_{n'}
\lambda(\vp,\vp';\varepsilon_{n}-\varepsilon_{n'})\,\mfG(\vp';\varepsilon_{n'})
\,,
\\
\hspace*{-2mm}
\Delta{\text{ep}}(\vp;\varepsilon_n)\ns 
=
\ns \int d\vp'\,T\sum_{n'}
\lambda(\vp,\vp';\varepsilon_{n}-\varepsilon_{n'})\,\mfF(\vp';\varepsilon_{n'})
\,,
\label{eq-Self-Energy-ep-off-diagonal}
\eer
where the phonon mediated electron-electron interaction,
\be\label{eq-lambda}
\lambda(\vp,\vp';\omega_m)
=
\sum_{\nu}
\vert g_{\nu}(\vp,\vp')\vert^2\,\cD_{\nu}(\vp-\vp',\omega_m)
\,,
\ee
is determined by the electron-phonon coupling, $g_{\nu}(\vp,\vp')$, and the equilibrium phonon propagator, $\cD_{\nu}(\vq,\omega_m)$, where $\omega_m = 2\pi T\,m$ are Boson Matsubara frequencies. Acoustic and optical phonons are labeled by a branch index $\nu$, and have energies $\omega_{\nu}(\vq)$ that disperse with momentum $\vq$ throughout the first Brillouin zone. Each branch contributes to the phonon-mediated interaction between electrons, with momenta $\vp$ and $\vp'$, with weights determined by the electron-phonon couplings $g_{\nu}(\vp,\vp')$. The spectral representation of the phonon propagator leads to a spectral representation of the electron-phonon interaction 
\be\label{eq-lambda-alpha2F}
\lambda(\vp,\vp';\omega_m)
=
\int_0^{\infty}d\omega'\,\alpha^2F(\vp,\vp';\omega')
\frac{2\,\omega'}{\omega'^2+\omega_m^2}
\,,
\ee
where 
\ber\label{eq-alpha2Fpp'}
\alpha^2F(\vp,\vp';\omega)=
N_f\int\ns d\varepsilon_{\vp}\ns\int\ns d\varepsilon_{\vp'}
\sum_{\nu}|g_{\nu}(\vp,\vp')|^2\,
\delta(\varepsilon_{\vp}-E_f)\,\delta(\varepsilon_{\vp'}-E_f)\,
\delta(\omega-\omega_{\nu}(\vp-\vp'))
\,,
\eer
is the angle-resolved electron-phonon spectral function. Equation \eqref{eq-alpha2Fpp'} is the generalization  of the Eliashberg's function, $\alpha^2F(\omega)$, for anisotropic electron-phonon interactions. The latter can be defined as the Fermi-surface-averaged spectral function, 
\be\label{eq-alpha2F-isotropic}
\alpha^2F(\omega)\equiv\int d\vp\int d\vp'\alpha^2F(\vp,\vp';\omega)
\,,
\ee
and is the spectral function that is often obtained from analysis of tunneling conductance data for strong-coupling superconductors~\citep{mcm65,wol80,arn80}.

%--------------------------------------------------------------------------------------

\subsubsection{Electronic Pairing Self-Energy}

The leading order contribution to the electron-electron self-energy is represented by the diagram in Fig.~\ref{fig-Sigma-Eliashberg}(c), which is generated by the renormalized Coulomb interaction, $V_{ee}(\vp,\vp';\varepsilon_n,\varepsilon_{n'})$, and which represents electron-electron scattering contributions to the self energy in both the particle-hole (Landau) and particle-particle (Cooper) channels.
The renormalized electron-electron interaction that defines the electron-electron self energy, $\whSigma_{ee}$, includes exchange. As a result $\whSigma_{ee}$ separates into spin-singlet and spin-triplet components in both the particle-hole (diagonal) and particle-particle (off-diagonal) self energies.
The internal line in Fig.~\ref{fig-Sigma-Eliashberg}(c)is the Nambu matrix propagator, $\whmfG$, representing intermediate particle-hole or Cooper pair excitations. Since both external and internal propagators are restricted to a low-energy shell near the Fermi surface, then to leading order in \sml, $V_{ee}$ can be evaluated with momenta and energies on the Fermi surface.\footnote{N.B. In cases where strong correlations drive the metallic ground state near to magnetic ordering the renormalized electron-electron interaction develops frequency dependence in the low-energy bandwidth, in which case retardation effects resulting from low-frequency magnetic fluctuations need to be included in the one-loop Fermionic self-energy.} 
Thus, for homogeneous equilibrium states the diagonal (Landau mean field) self energy vanishes. The off-diagonal contribution to the electronic self energy is non-zero below $T_c$.

The renormalized electron-electron interaction in the Cooper channel can be expressed in terms of dimensionless interaction potentials,
\ber\label{eq-Vee-pp}
\hspace*{-3mm}
N_f[V_{\text{ee}}]_{\alpha\beta;\gamma\rho}(\vp,\vp')
=
\mu^{(s)}(\vp,\vp')(i\sigma_y)_{\alpha\beta}(i\sigma_y)_{\gamma\rho}
+
\mu^{(t)}(\vp,\vp') 
(i\sigma_y\vec\sigma)_{\alpha\beta}\cdot(i\vec\sigma\sigma_y)_{\gamma\rho}
\,,
\eer
that separate into spin-singlet (total spin $S=0$) and spin-triplet ($S=1$) channels, labeled by superscripts, $(s,t)$, with corresponding interactions between pairs of quasiparticles with zero total momentum, $\mu^{(s)}(\vp,\vp')$ and $\mu^{(t)}(\vp,\vp')$, respectively. 
For Nb the triplet pairing channel is at best sub-dominant to the singlet channel. Thus, we ignore triplet pairing correlations for the homogeneous equilibrium state of Nb.  
As a result the electron-electron anomalous propagator and off-diagonal self energy have the spin-singlet form, 
\be\label{eq-anomalous_propagator}
\hspace*{-2.5mm}
\whmfF
=
\begin{pmatrix}
0 & \mfF\,i\sigma_y
\cr 
\underline{\mfF}\,i\sigma_y & 0 
\end{pmatrix}
\,,
\ee
where $\mfF$ is the spin-singlet Cooper pair propagator. The corresponding electronic contribution to the off-diagonal pairing self energy decomposes similarly,
\be\label{eq-off-diagonal-self-energy}
\hspace*{-2.5mm}
\whDel_{\text{ee}}
=
\begin{pmatrix}
0 & \Del_{\text{ee}}\,i\sigma_y
\cr 
\uDel_{\text{ee}}\,i\sigma_y & 0 
\end{pmatrix}
\,.
\ee
In the absence of retardation resulting from the coupling to long-lived collective excitations, e.g. spin-fluctuations, then to leading order in \sml\ we can evaluate the renormalized electron-electron interaction for momenta and energies on the Fermi surface, in which case we obtain the off-diagonal pairing self energy generated by the renormalized electron-electron interaction in the singlet channel,
\ber
\Del_{\text{ee}}(\vp)&=&
-\int d\vp'\,
\mu^{(s)}(\vp,\vp')
T\ns\sum_{n}^{|\varepsilon_{n}|\le\omega_c}\,
\mfF(\vp';\varepsilon_n)
\,.
\label{eq-Self-Energy-ee-off-diagonal-singlet}
\eer
For electron-phonon-mediated superconductors like Nb, the Cooper instability is in the ``conventional'' spin-singlet, $A_{1g}$ channel. Thus, we need retain only the spin singlet, $A_{1g}$ component of $\mu^{(s)}(\vp,\vp')$. The corresponding renormalized electron-electron interaction is repulsive and competes with the attractive electron-phonon mediated pairing interaction, suppressing the instability temperature to superconductivity.

In what follows we neglect the angular dependence of the renormalized electron-electron interaction, in which case $\mu^{(s)}(\vp,\vp')\rightarrow\mu$, the isotropic average of the static screened Coulomb interaction.
Accurate calculation of Coulomb interaction is beyond DFT, but is estimated from the static screened Coulomb interaction for an electron-ion plasma, defined here as $\mu=N_f V_{ee}$.
The cutoff that regulates the electron-electron contribution to the gap equation is $\Omega\sim E_f\gg\hbar\omega_D$, where $E_f$ is the Fermi energy and $\omega_D$ is the Debye frequency.
However, the cutoff, $\omega_c$, that we introduce to regulate the electron-phonon contribution to the gap equation is $\omega_D < \omega_c \ll \Omega$. 
In Sec.~\ref{sec-Isotropic_Eliashberg_Theory} we describe the procedure used to determine the low energy cutoff $\omega_c$, which includes renormalization of the Coulomb interaction, $\mu\rightarrow\mu^{\star}$, such that we obtain a single gap equation for $\Delta$ with the low-energy cutoff $\omega_c$.

The spatially homogeneous solution to Eqs.~\eqref{eq-Eilenberger_Transport} and \eqref{eq-Eilenberger_Normalization} for the Nambu matrix propagator reduce to 
\be
\whmfG(\vp;\varepsilon_n) 
=
-\pi\frac{\tilde\varepsilon_n(\vp;\varepsilon_n) \whtauz 
         -\tilde\Delta(\vp;\varepsilon_n)\,i\sigma_y\,\whtaux}
{\sqrt{\tilde\varepsilon_n(\vp;\varepsilon_n)^2 + |\tilde\Delta(\vp;\varepsilon_n)|^2}}
\,,
\ee
with the renormalized Matsubara energy and pairing self energy defined by
$i\tilde\varepsilon_n \equiv i\varepsilon_n - \Sigma_{\text{ep}}$ and $\tilde\Delta \equiv \Delta_{\text{ep}} + \Delta_{\text{ee}}$. 
Evaluating Eqs.~\eqref{eq-Self-Energy-ep-diagonal},~\eqref{eq-Self-Energy-ep-off-diagonal} and~\eqref{eq-Self-Energy-ee-off-diagonal-singlet} with the corresponding propagators, and defining 
$Z(\vp;\varepsilon_n)\equiv\tilde\varepsilon_n/\varepsilon_n$, 
and 
$\Delta(\vp;\varepsilon_n)\equiv\tilde\Delta(\vp;\varepsilon_n)/Z(\vp;\varepsilon_n)$, gives the Eliashberg's equations including the renormalized Coulomb interaction~\citep{mor62,sca66,rai86},
%
%------------------------------------------------------------------------------
\ber
Z(\vp;\varepsilon_n) 
&=& 
1 + \frac{1}{\varepsilon_n} 
\pi T\sum_{n'}\,
\int d\vp'\,
\lambda(\vp,\vp';\varepsilon_n-\varepsilon_{n'})
\frac{\varepsilon_{n'}}{\sqrt{\varepsilon_{n'}^2 + |\Delta(\vp';\varepsilon_{n'})|^2}}
\,,
\label{eq-Eliashberg-Z}
\\
Z(\vp;\varepsilon_n)\,\Delta(\vp;\varepsilon_n) 
&=& 
\pi T\sum_{n'}\,
\int d\vp'\,
\left[\lambda(\vp,\vp';\varepsilon_n-\varepsilon_{n'})-\mu^{\star}\right]
\frac{\Delta(\vp';\varepsilon_{n'})}
     {\sqrt{\varepsilon_{n'}^2 + |\Delta(\vp';\varepsilon_{n'})|^2}}
\,.
\label{eq-Eliashberg-Delta}
\eer
%------------------------------------------------------------------------------
%
If the anisotropy of the pairing self energy is negligible then we can simplify Eqs.~\eqref{eq-Eliashberg-Z}-\eqref{eq-Eliashberg-Delta} by averaging the electron-phonon spectral function to obtain Eq.~\eqref{eq-alpha2F-isotropic}, and the reduction of Eqs.~\eqref{eq-Eliashberg-Z} and \eqref{eq-Eliashberg-Delta} to the simpler set of integral-sum equations,
\ber
Z(\varepsilon_n) 
&=& 
1 + \frac{1}{\varepsilon_n}\pi T \sum_{n'}  
\lambda(\varepsilon_{n}-\varepsilon_{n'})\times 
\frac{\varepsilon_{n'}}{\sqrt{\varepsilon_{n'}^2+|\Delta{(\varepsilon_{n'})|^2} }}
\label{eq-Eliashberg-Z-isotropic}
\,,
\\ 
Z(\varepsilon_n)\,\Delta(\varepsilon_{n}) 
&=& 
\pi T\sum_{n'}\left[\lambda(\varepsilon_n-\varepsilon_{n'})-\mu^{\star}\right]
\times
\frac{\Delta(\varepsilon_{n'})}{\sqrt{\varepsilon_{n'}^2+|\Delta(\varepsilon_{n'})|^2}} 
\label{eq-Eliashberg-Delta-isotropic}
\,.
\end{eqnarray}
where the electron-phonon coupling function averaged over the Fermi-surface is
\ber
\lambda(\omega_m)
=
\int d\vp\,\int d\vp'\,\lambda(\vp,\vp';\omega_m)
=
2\int_0^{\infty}d\omega'\,\alpha^2F(\omega')\,\frac{\omega'}{\omega'^2+\omega_m^2}
\,,
\label{eq-lambda-isotropic}
\eer
with $\alpha^2F(\omega)$ defined by Eqs.~\eqref{eq-alpha2F-isotropic} and \eqref{eq-alpha2Fpp'}.
The isotropic Eliashberg's equations are the correct limiting equations for strong-coupling superconductors in the dirty limit where diffusive motion of electrons averages the electron-phonon interaction over the Fermi surface.
In Sec.~\ref{sec-Tc-disorder} we examine the effect of scattering by a random impurity potential and calculate $T_c$ in anisotropic superconducting Niobium as a function of the quasiparticle-impurity scattering rate, $1/\tau$. 

%------------------------------------------------------------------------------

\section{Electronic Structure}\label{sec-DFT}

The electronic structure and superconducting state of Nb has been the subject of considerable theoretical, computational and experimental investigation~\citep{tow62,mcm68,mat70,bos76,but77,pow77,wol80,daa80,sav96,sau20}.
Accurate results for the energy levels and dispersion relations for electrons and phonons, as well as the interaction between electrons and phonons, are essential for calculating the superconducting properties of Nb~\citep{rai86}.
We first obtain the electronic band structure and phonon dispersion relations for Nb using Quantum Espresso (QE) which is an integrated suite of open-source computer codes for electronic-structure calculations and materials modeling at the atomic scale. QE is based on density functional theory, plane waves, and pseudo-potentials~\citep{gia09,gia17}.
From the electronic band structure data we then construct the Fermi surface and calculate the Fermi velocity at each point on the Fermi surface. The anisotropy of the Fermi velocity plays a central role in determining the anisotropy of the upper critical field of Nb~\citep{sei78,ara04}.
We obtain the electron-phonon interaction and phonon spectral function, and use Eliashberg theory to calculate the superconducting order parameter (``gap function'') as a function of momentum on the Fermi surface and for energies within the phonon bandwidth of attraction.

%------------------------------------------------------------------------------
\begin{figure}[t!]
\centering\includegraphics[width=0.75\textwidth]{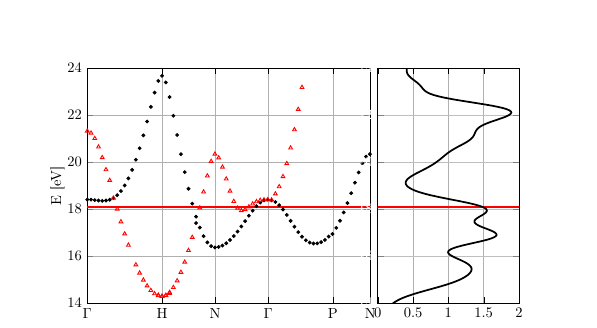}
\caption{
The red line is the Fermi energy at $18.096\, eV$. Shown are the 2 electronic bands of Niobium in phonon bandwidth that cross the Fermi energy. 
The right panel shows the electronic density of states in units of $eV^{-1}$.
}
\label{fig-Electronic_Bands-Fermi_Level}
\end{figure}
%------------------------------------------------------------------------------

Gap anisotropy plays a key role in pair-breaking processes associated with impurity and boundary scattering. In particular, the combination of branch conversion scattering induced by impurity scattering leads to suppression of the superconducting transition temperature.
We compute the suppression of $T_c$ using self-consistent T-matrix theory for quasiparticle-impurity scattering for the broad class of anisotropic superconductors and use the result to predict the suppression of $T_c$ of Niobium doped with non-magnetic impurities. We compare our results with reports of the suppression of $T_c$ for N-doped Nb SRF cavities as well as disordered Nb films in Sec.~\ref{sec-Tc-disorder}. 

Bulk single crystals of Niobium have BCC lattice structure with lattice constant $a=3.3\,\mbox{\AA}$ and atomic weight of $M=92.906$. The electron configuration of the Niobium atom is $[Kr]4d^45s^1$, which generates 24 electronic bands. 
Superconductivity develops from pairing of electrons and holes in a narrow band of energies near the Fermi surface. 
%
%------------------------------------------------------------------------------
\begin{figure}[t!]
\begin{minipage}{\textwidth}
\includegraphics[width=0.5\textwidth]{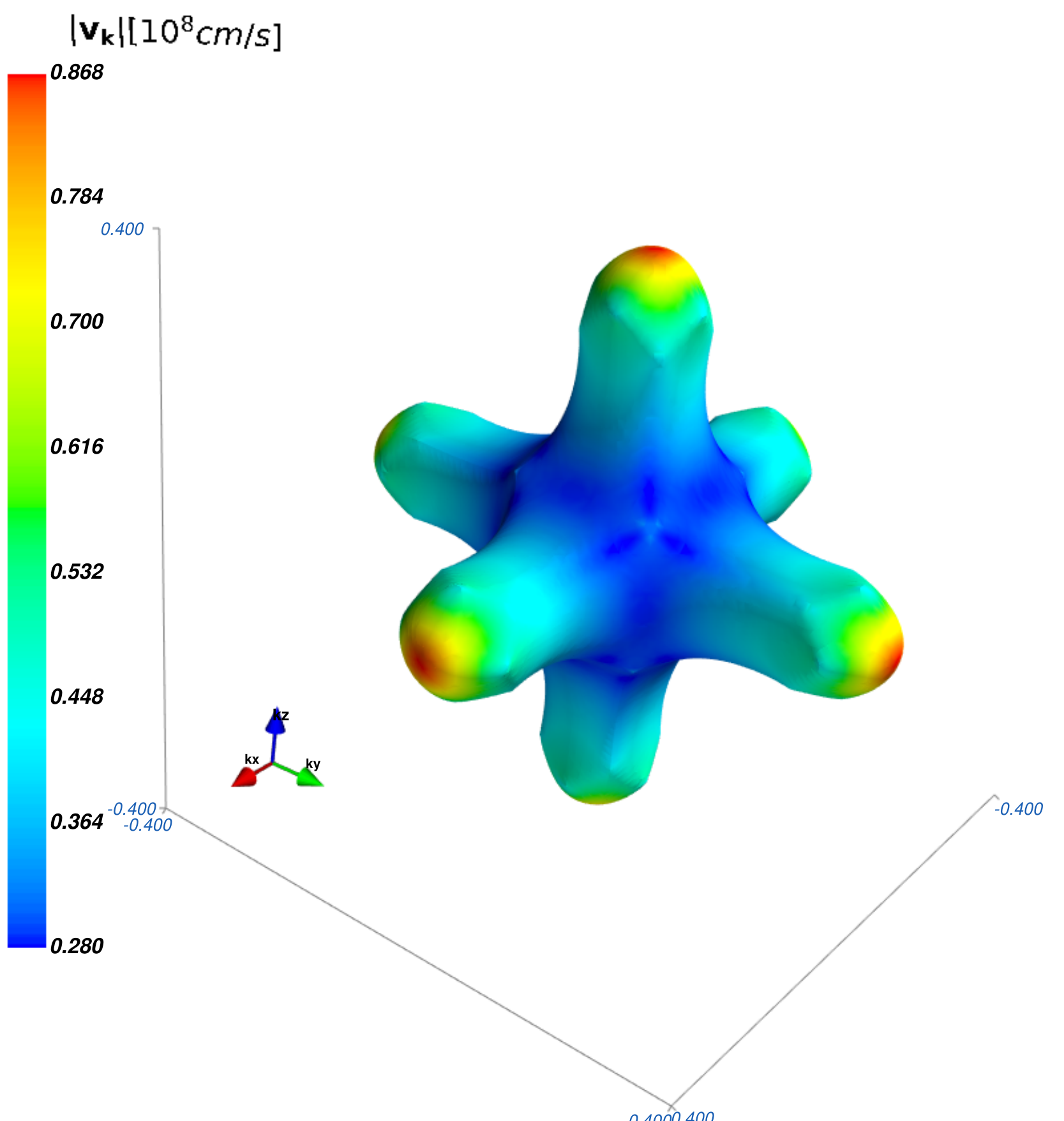}\includegraphics[width=0.5\textwidth]{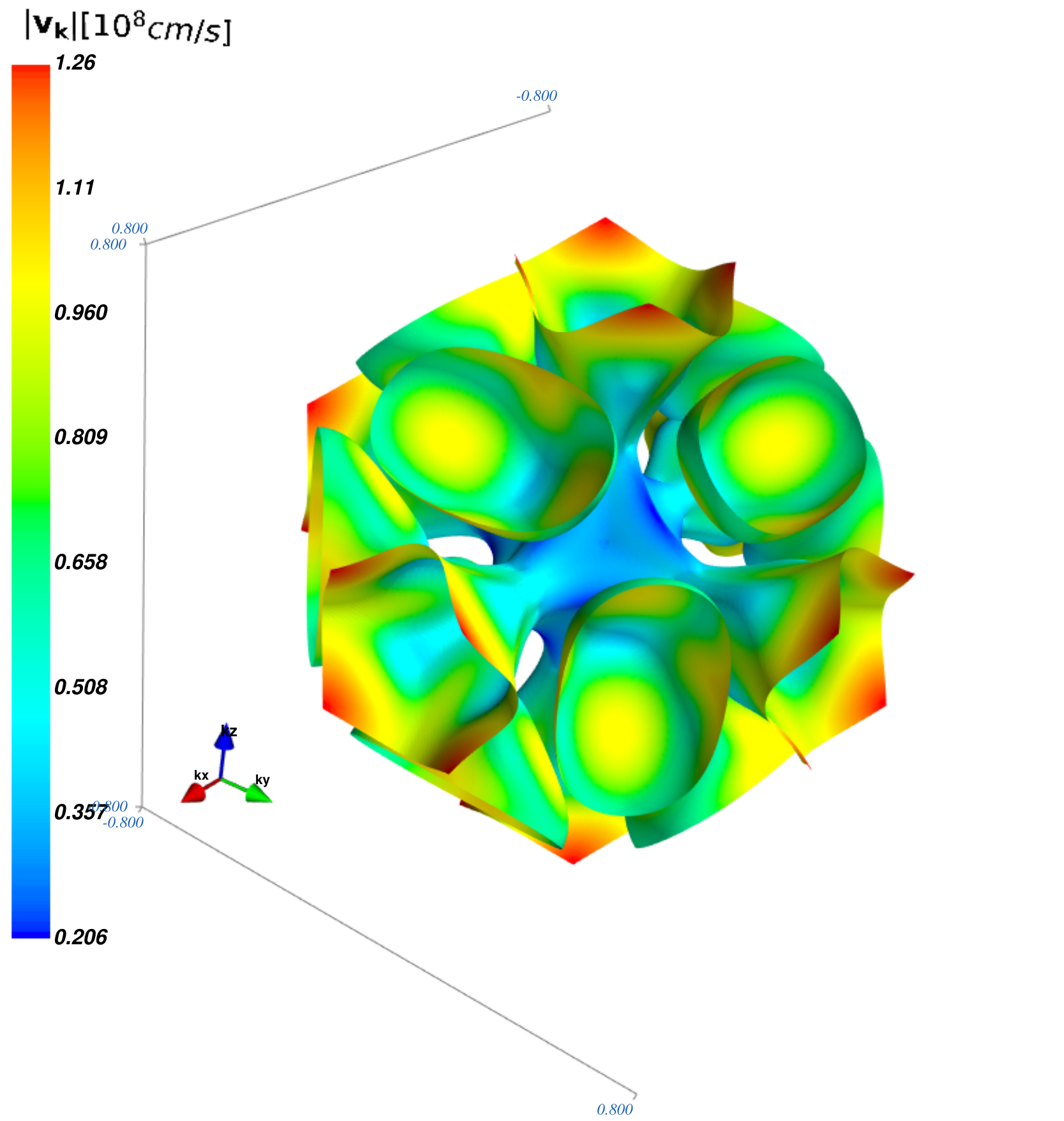}
\caption{The anisotropy of the Fermi velocity shown on the sheets of the Fermi surface defined in the FBZ.}
\label{fig-Fermi_velocity}
\end{minipage}
\end{figure}
%------------------------------------------------------------------------------
%
An accurate calculation of the superconducting order parameter (``gap'') requires numerical integration over fine grids in momentum space for the energy levels of electrons and phonons. A direct calculation of the electron and phonon band structures, as well as the electron-phonon coupling and spectral function, on fine grids is computationally demanding.
A more efficient approach was developed by the authors of Ref.~\citep{piz20,mar97,sou01}. The method is to calculate the electronic bands on a coarse grid in momentum space, but over a wider bandwidth around the Fermi surface, Fourier transform to coordinate space and find maximally localized Wannier functions. The wider energy bandwidth results in more accurate Wannier functions. Once determined one can Fourier transform back to a finer grid in momentum space.

Figure~\ref{fig-Electronic_Bands-Fermi_Level} shows the two lowest energy electronic bands of Nb that cross the Fermi energy for the course grid of $k=98^3$ points in the First Brillouin Zone (FBZ) calculated using QE. 
The right panel of Fig.~\ref{fig-Electronic_Bands-Fermi_Level} shows the electronic density of states (DOS) for the same low-energy bandwidth.
Using the band structure calculated for a uniform $k$-grid and the Fermi energy of $E_f=18.096\,\mbox{eV}$, we construct the Fermi surface using the \textit{marching cube algorithm} to identify the Fermi surface in momentum space~\citep{lor87}. The result is shown in Fig.~\ref{fig-Fermi_velocity} for two sheets of the Fermi surface in the FBZ.
The left panel of Fig.~\ref{fig-Fermi_velocity} shows the Fermi surface sheet referred to as the ``jack'' centered at the $\Gamma$ point, while the right panel shows the open Fermi sheet referred to as the``jungle gym'', also centered at the $\Gamma$ point, and the ``ellipsoids'' centered on the $N$ points.

From the band dispersions near the Fermi energy we calculated the group velocity, $\vv_{\vp}=\grad_{\vp}E_{\vp}$, evaluated at the Fermi energy, i.e.\ the Fermi velocity, at each point on the Fermi surface. The color map shown in Fig.~\ref{fig-Fermi_velocity} indicates the magnitude of Fermi velocity at each point on the Fermi surface. 
There is substantial anisotropy of the Fermi velocity with a maximum velocity of $v_f^{\text{max}}=1.26\times 10^8\,\mbox{cm/s}$, a minimum of $v_f^{\text{min}}=0.28\times 10^8\,\mbox{cm/s}$. Table~\ref{table-Fermi_velocity} summarizes the average velocity, $\bar{v}_f$, \emph{rms} average velocity, $v_f^{\text{rms}}$, and the standard deviation for the distribution of Fermi velocity, $\sigma_{v}=\sqrt{\mathcal{A}_{v}}\,v_f^{\text{rms}}$, where    
$\mathcal{A}_{v}\equiv 1-\frac{\langle|\vv_{\vp}|\rangle^2}{\langle|\vv_{\vp}|^2\rangle}$,
and $\langle\ldots\rangle\equiv\int d\vp (\ldots)$.
Table~\ref{table-Fermi_velocity} summarizes results for the anisotropy of the Fermi velocity on the two sheets of the Fermi surface in the FBZ. The contribution to the DOS from each sheet is also shown in Table~\ref{table-Fermi_velocity}.
The data for the anisotropy of Fermi velocity data is important for prediction and analysis of the anisotropy of the upper critical field of Niobium~\citep{ara04}.

%------------------------------------------------------------------------------
% Table: Fermi velocity
%
\begin{table}[t]
\begin{center}
\begin{tabular}{|c|c|c|c|c|c|}
\hline
Sheet
&
$\bar{v}_f\equiv\langle|\vv_{\vp}|\rangle$ 
& 
$v_{f}^{\text{rms}}\equiv\sqrt{\langle |\vv_{\vp}|^2\rangle}$
&
$\sigma_{v}$
&
$\mathcal{A}_{v}$ 
&
$\nu(E_f)$ 
\\
\hline
1 & 0.419 & 0.437 & 0.122 & 0.078 & 0.183\\
\hline
2 & 0.732 & 0.762 & 0.214 & 0.079 & 1.149\\
\hline
\end{tabular}
\end{center}
\caption{The mean Fermi velocity, the \mbox{rms} Fermi velocity, and the standard deviation of the distribution of Fermi velocities, in units of $10^{8}cm/s$. 
In addition, the dimensionless anisotropy parameter, $\cA_v$, and the contribution to the DOS at the Fermi energy in units of $\mbox{eV}^{-1}$.}
\label{table-Fermi_velocity}
\end{table}
%-----------------------------------------------------------------------------

\subsection{Phonon Band Structure}

\begin{figure}[t!]
\centering\includegraphics[width=0.75\textwidth]{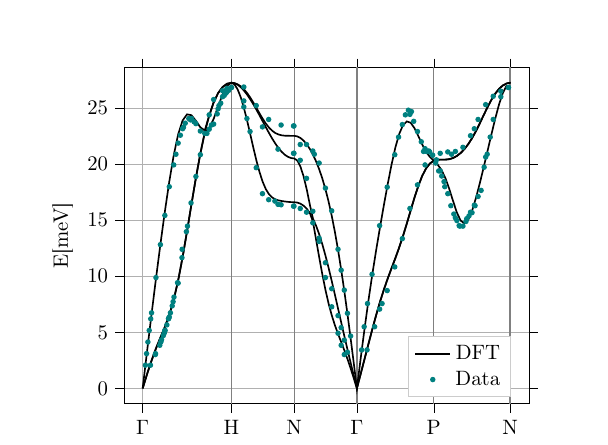}
\caption{Calculated phonon energies (solid-lines) for Nb based on QE along various directions in the FBZ. The green data points are the phonon energies obtained from inelastic neutron scattering~\citep{pow77}}.
\label{fig-Phonon_Dispersion}
\end{figure}

The phonon band structure is calculated based on electronic structure calculations and the Born-Oppenheimer approximation. For the purposes of calculating the phonon energy levels and bandstructure this allows one to decouple the dynamics of the electronic subsystem from the lattice dynamics. 
Thus, the ground state energy of the electronic system is calculated for the fixed ionic positions. The resulting total energy of the electronic system $E_{\mathrm{el}}(\mathbf{R}_1, \mathbf{R}_2, \dots, \mathbf{R}_N)$, serves as a potential energy function for the ionic Hamiltonian. By displacing the atoms by small amounts, $\{\mathbf{u}_i, i=1,..,N\}$, relative to the Bravais lattice sites, the electronic ground state for displaced ions is calculated. The ionic lattice energy is then expanded in displacements of the ions relative to their equilibrium BCC lattice configuration. The first derivatives of the energy functional vanish, and the set of second derivatives provides a matrix of interactions between displaced ions. The Fourier transform of this matrix with respect relative displacements gives the \emph{dynamical matrix} whose eigenvalues determine the phonon energies, $\omega_{\nu}(\vq)$, where $\nu$ is the phonon branch index and $\vq$ is the phonon wavevector. 
The dynamical matrix is calculated using the QE code for a discrete grid of wavevectors $\vq$ belonging to the unit cell in reciprocal space~\citep{deg95}. Since the dynamical matrix is a smooth function of $\vq$, it is  usually sufficient to evaluate the matrix on a sparse grid in reciprocal space, then perform a discrete Fourier transform to the position space, restrict the inter-atomic forces to a few lattice spacings, and finally transform back to momentum space to obtain the dynamical matrix on much finer grid in reciprocal space. The eigenvalues of the resulting dynamical matrix generate the phonon dispersion relations evaluated on the dense grid in reciprocal space~\citep{deg95}. 
Figure \ref{fig-Phonon_Dispersion} shows the results for our calculation of the phonon modes in comparison with the modes obtained from inelastic neutron scattering~\citep{pow77}.

%------------------------------------------------------------------------------
%------------------------------------------------------------------------------

\subsection{Electron-Phonon Coupling}

The retarded electron-phonon interaction defined by Eqs.~\eqref{eq-lambda-alpha2F} and \eqref{eq-alpha2Fpp'} depends on the electron-phonon matrix element, $g_{\nu}(\vp,\vp')$.  
Thus, the transition temperature and superconducting order parameter depend on an accurate determination of the electron-phonon matrix element and phonon density of states, both of which are anisotropic.
The matrix element for the scattering of an electron with momentum $\vp$ to a state with momentum $\vp'$ by a phonon of branch $\nu$ and momentum $\vq=\vp'-\vp$ based on perturbation theory in the ionic displacement is~\citep{pon16},
\begin{eqnarray}
g_{\nu}(\vp;\vp')
\ns&=&\ns
\frac{1}{\sqrt{2\omega_{\nu}(\vq)}}
\bra{\psi_{\vp'}}\partial^{\nu}_{\vq}V\ket{\psi_{\vp'}}
,\quad
\\
\mbox{where}\quad
\partial^{\nu}_{\vq}V 
&\equiv&
\hat{\ve}^{\nu}_{\vq}\cdot\boldsymbol{\nabla}_{\vR} V
\,\mbox{and}\,\vq = \vp'-\vp
\,.\quad
\label{eq-e_ph-derivative}
\end{eqnarray}

The self-consistent electron-nucleus interaction potential, $V(\vr-\vR)$, is calculated for small ion displacements, $\vu=\vR-\vR_{\text{BL}}$, where $\vR_{\text{BL}}$ is an equilibrium Bravais lattice site. The directional derivative in Eq.~\eqref{eq-e_ph-derivative} is defined by the polarization vector of the phonon, $\hat\ve^{\nu}_{\vq}$.
The nuclear mass  enters via the phonon frequencies, and $\ket{\psi_{\vp}}$ is the Kohn-Sham electronic orbital for momentum $\vp$.

\section{Results}

We use EPW, which is an integral part of QE, to compute the electron-phonon matrix elements. The calculation of these quantities requires dense grids in reciprocal space. To achieve such dense grids it is efficient to Fourier transform the Kohn-Sham orbitals to position space, construct optimally localized Wannier functions, then Fourier transformation back to a obtain a dense grid in momentum space~\citep{giu07,pon16,mar13}.

From the electron-phonon matrix elements and the phonon spectrum the electron-phonon spectral function and pairing interaction function are computed using Eqs.~\eqref{eq-alpha2Fpp'} and \eqref{eq-lambda-alpha2F}. The corresponding Fermi-surface averaged quantities, Eqs.~\eqref{eq-alpha2F-isotropic} and \eqref{eq-lambda-isotropic}, are calculated by averaging over the Fermi surface.

%------------------------------------------------------------------------------
\begin{figure}[t!]
\centering\includegraphics[width=0.75\textwidth]{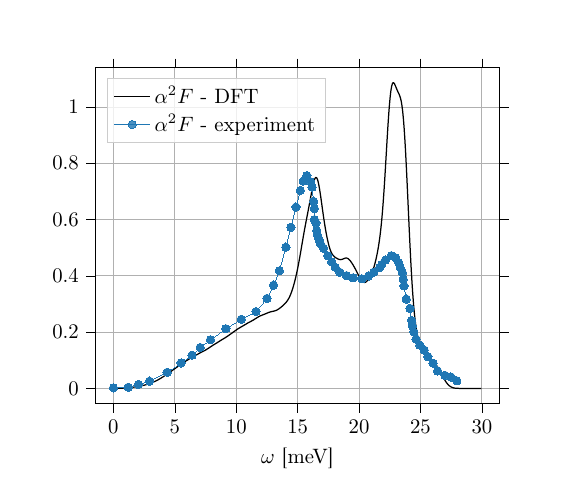}
\caption{Calculated Fermi-surface averaged $\alpha^2F(\omega)$ as a function of phonon frequency (solid line). Experimental data (blue dots and line) derived from tunneling spectroscopy~\citep{arn80}.
}
\label{fig-alpha2F-isotropic}
\end{figure}
%------------------------------------------------------------------------------

\subsection{Isotropic Eliashberg Theory}\label{sec-Isotropic_Eliashberg_Theory}

The isotropic electron-phonon spectral function, $\alpha^2F(\omega)$, is related to the differential conductance for NIS tunneling into strong-coupling superconductors~\citep{bos76,but77,rob78,wol80,daa80}. For comparison we show the results reported in Ref.~\citep{arn80} for Nb in Fig.~\ref{fig-alpha2F-isotropic} in comparison with our result for the calculated spectral function using EPW.
The low frequency values are in reasonable agreement with the data from tunneling experiments, however there are deviations for the phonon modes near the zone boundary, particularly the high-frequency longitudinal phonon near $23\,\mbox{meV}$. The transverse phonon peak is calculated to be slightly higher in frequency than the experimental peak at $15.75\,\mbox{meV}$.
Previous ab-initio calculations also report higher spectral weight for the longitudinal phonon peak than that obtained from tunneling spectroscopy~\citep{but79,sav96}. As other authors have noted a determination of $\alpha^2F(\omega)$ from tunneling spectroscopy is subject to a number of interface effects that may complicate an accurate inversion of the tunneling data for the electron-phonon coupling function~\citep{wol80,arn80,daa80,sav96}.

%------------------------------------------------------------------------------
\begin{figure}[t!]
\centering\includegraphics[width=0.75\textwidth]{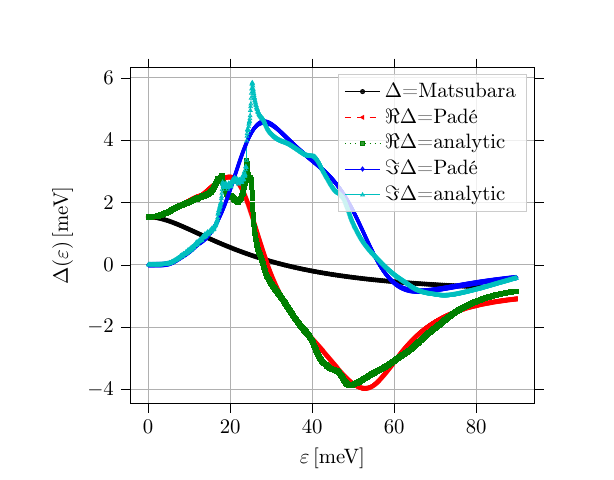}
\caption{The superconducting gap $\Delta(\varepsilon)$ at $T=1.0\, K$ as a function of Matsubara energy $\varepsilon_n$ (black circles). The real and imaginary parts of the gap function calculated from the gap equation defined on the real energy axis are labelled `analytic'. The same functions obtained by numerical continuation of the Matsubara gap function using Pad\'e approximates are labelled `Pad\'e'.}
\label{fig-Delta_vs_epsilon}
\end{figure}
%------------------------------------------------------------------------------

From our calculated result for $\alpha^2F(\omega)$ shown in Fig.~\ref{fig-alpha2F-isotropic} the average electron-phonon coupling is calculated to be
\be
\lambda(0)=2\int_0^{\infty}d\omega\frac{\alpha^2F(\omega)}{\omega}\approx 1.057
\,.
\ee
This result compares with the value of $1.14$ obtained in Ref.~\citep{sco70} based on de Hass-van Alphen measurements, as well as tunneling spectroscopy, $1.04$ from Ref.~\citep{wolf12} and $0.98$ from Ref.~\citep{rob78}.

The Fermi-surface averaged electron-phonon spectral function $\alpha^2F(\omega)$ is used to calculate the superconducting order parameter and $T_c$ from the isotropic Eliashberg equations, Eqs. \eqref{eq-Eliashberg-Z-isotropic}-\eqref{eq-Eliashberg-Delta-isotropic}, as a function of energy within the phonon bandwidth and as a function of temperature. 
For pure Nb with a transition temperature of $T_c=9.33\,\mbox{K}$ and momentum grids of $k=q=51^3$ we obtain a renormalized Coulomb interaction of $\mu^{\star}=0.218$ for a cutoff of $\omega_c=3\omega_D$.\footnote{The theoretical value might be slightly smaller as $\mu^{\star}$ tends to decrease for finer momentum grids. In particular, for a coarser grid with $k=q=36^3$ we obtain $\mu^{\star}=0.260$ with no change in $\Delta(T)$.} 
This value of $\mu^{\star}$ is close to the experimentally determined value of $0.24$ based on analysis of de Hass-van Alphen data~\citep{sco70}. 

Figure~\ref{fig-Delta_vs_epsilon} shows the calculated gap function for $T=1\,\mbox{K}$ as a function of Matsubara frequency (smooth black curve), as well as both real and imaginary components of $\Delta^{\text{R}}(\varepsilon)$ obtained by analytic continuation to the real frequency axis. The QE code calculates $\Delta^{\text{R}}(\varepsilon)$ from the gap equation analytically continued to the real axis, as well as by numerical continuation using Pad\'e approximates of $\Delta(\varepsilon_n)$ defined on Matsubara frequencies. The results are shown in Fig.~\ref{fig-Delta_vs_epsilon}.

%------------------------------------------------------------------------------
\begin{figure}[h]
\begin{minipage}{0.475\textwidth}
\centering\includegraphics[width=\textwidth]{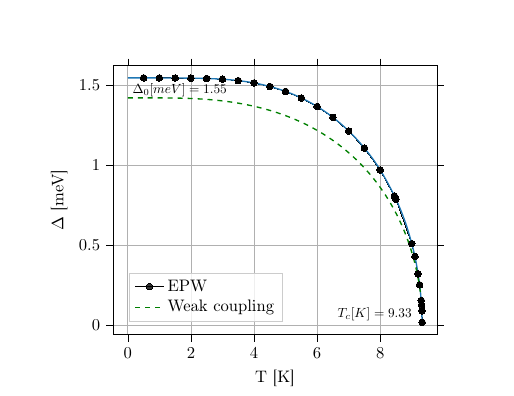}
\caption{The superconducting gap as a function of temperature (dots and solid line) for momentum grids of $k=51^3,q=51^3$ and $\mu^{\star}=0.218$. The dashed line is the result for the gap calculated using the weak-coupling BCS theory.}
\label{fig-Delta_vs_T}
\end{minipage}
\hspace*{1cm}
\begin{minipage}{0.475\textwidth}
\centering\includegraphics[width=\textwidth]{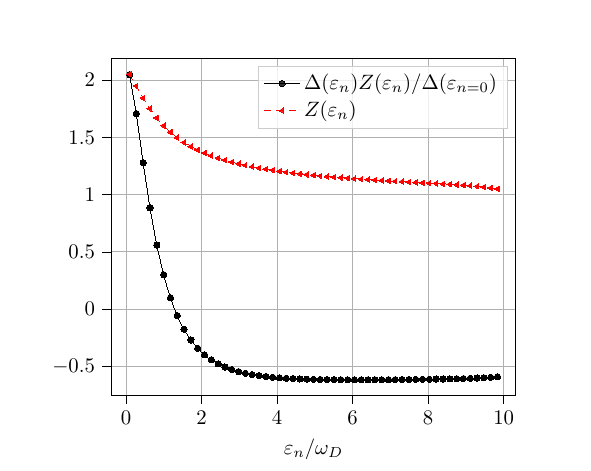}
\caption{The diagonal and off-diagonal self-energy function, derived by solving Eliashberg equations, as a function of $\varepsilon_n/\omega_D$ where $\varepsilon_n=(2n+1)\pi T$ is Matsubara frequency.}
\label{fig-Z-vs-wc}
\end{minipage}
\end{figure}
%------------------------------------------------------------------------------

For strong-coupling superconductors the gap obtained using tunneling conductance spectroscopy is identified with $\Delta(\varepsilon_n\rightarrow\pi T)$. Figure~\ref{fig-Delta_vs_T} shows the tunneling gap as a function of the temperature. The value of the gap at zero temperature is $\Delta_0=1.55\,\mbox{meV}$ and a transition temperature of $T_c= 9.33\,\mbox{K}$. This result is also in reasonable agreement with the value of $1.5\,\mbox{meV}$ reported by several independent studies based on tunneling spectroscopy~\citep{she61,tow62,arn80}. 
This corresponds to a modest enhancement of the zero-temperature gap compared to the weak-coupling BCS prediction of $\Delta_0^{\text{wc}}=1.78\,\kb\,T_c\approx 1.43\,\mbox{meV}$. For comparison we also show in Fig.~\ref{fig-Delta_vs_T} the weak-coupling BCS prediction for the gap for the same $T_c$.

The low-energy cutoff, $\omega_c$, is chosen such that the solution of the Eliashberg equation for $\tilde\Delta(\varepsilon_n)\equiv Z(\varepsilon_n)\Delta(\varepsilon_n)$ in Eqs.~\eqref{eq-Eliashberg-Z-isotropic} and~\eqref{eq-Eliashberg-Delta-isotropic} becomes independent of the cutoff.
In Fig.\ref{fig-Z-vs-wc} we plot the renormalization factors for the Matsubara energies, 
$Z(\varepsilon_n)\equiv\tilde\varepsilon_n/\varepsilon_n$, 
and the renormalized gap ratio, 
$\Delta(\varepsilon_n)Z(\varepsilon_n)/\Delta(\varepsilon_{n=0})$,
as a function of $\varepsilon_n/\omega_D$.
Both ratios saturate for $\varepsilon_n\gtrsim 3\omega_D$. Thus, we can choose the lower cutoff as $\omega_c=3\omega_D$~\citep{rai86}.
As a check we have also carried out calculations with $\omega_c= 10 \omega_D$ and find no significant change in the value of the zero temperature gap.
However, $\mu^{\star}$ includes the reduction in the Coulomb repulsion due to retardation of the electron-phonon mediated interaction over timescales of order the inverse of phonon bandwidth $1/\omega_{D}$ compared to the nearly instantaneous Coulomb repulsion that operates on the much shorter timescale of $1/\Omega$~\citep{mor62}. The result is the re-normalized Coulomb interaction $\mu^{\star}$ given by $1/\mu^{\star}=1/\mu + \ln(\Omega/\omega_c)$, which depends weakly on $\omega_c$.
For the higher cutoff, $\mu^{\star}(10\omega_D)\simeq 0.253$ compared to $\mu^{\star}(3\omega_D)\simeq 0.218$. Compared to the gap calculated with $\omega_c=3\omega_D$ the higher cutoff leaves the average gap function unchanged. However, there is a slight change in the anisotropy of the gap corresponding to $\mathcal{A}=0.032$ for $\omega_c=10\omega_D$ compared to $\mathcal{A}=0.037$ for $\omega_c=3\omega_D$.

\subsection{Anisotropic Eliashberg Theory}

Anisotropy of the electron-phonon coupling function, $\lambda(\vp,\vp';\omega_m)$, and thus the superconducting gap function, $\Delta(\vp;\varepsilon_n)$, is an important and widely discussed topic~\citep{she65,vic68,sel73,bos73,but76,cra79,sei78,but80,pet77}.
Our analysis implies measurable gap anisotropy for pure Nb and shows that the electron-phonon matrix element and phonon density of states are anisotropic functions of the momenta on the Fermi surface. The anisotropy of $\alpha^2F(\vp,\vp';\omega)$ generates an anisotropic pairing self energy, $\Delta(\vp;\varepsilon_n)$, obtained using EPW as the solution of the anisotropic Eliashberg equations, \eqref{eq-Eliashberg-Z} and \eqref{eq-Eliashberg-Delta}.
The EPW code calculates $\Delta^{\text{R}}(\vp;\varepsilon)$ in an energy shell of order $\delta\varepsilon=1\,\mbox{eV}$ around the Fermi surface. This rather thick shell is require in order to obtain accurate results for the self-energies. We then determine the Fermi momentum and $\Delta(\vp;\varepsilon_n)$ on the Fermi surface by linear interpolation.

%------------------------------------------------------------------------------
\begin{figure}[t!]
\begin{minipage}{\textwidth}
\centering\includegraphics[width=0.5\textwidth]{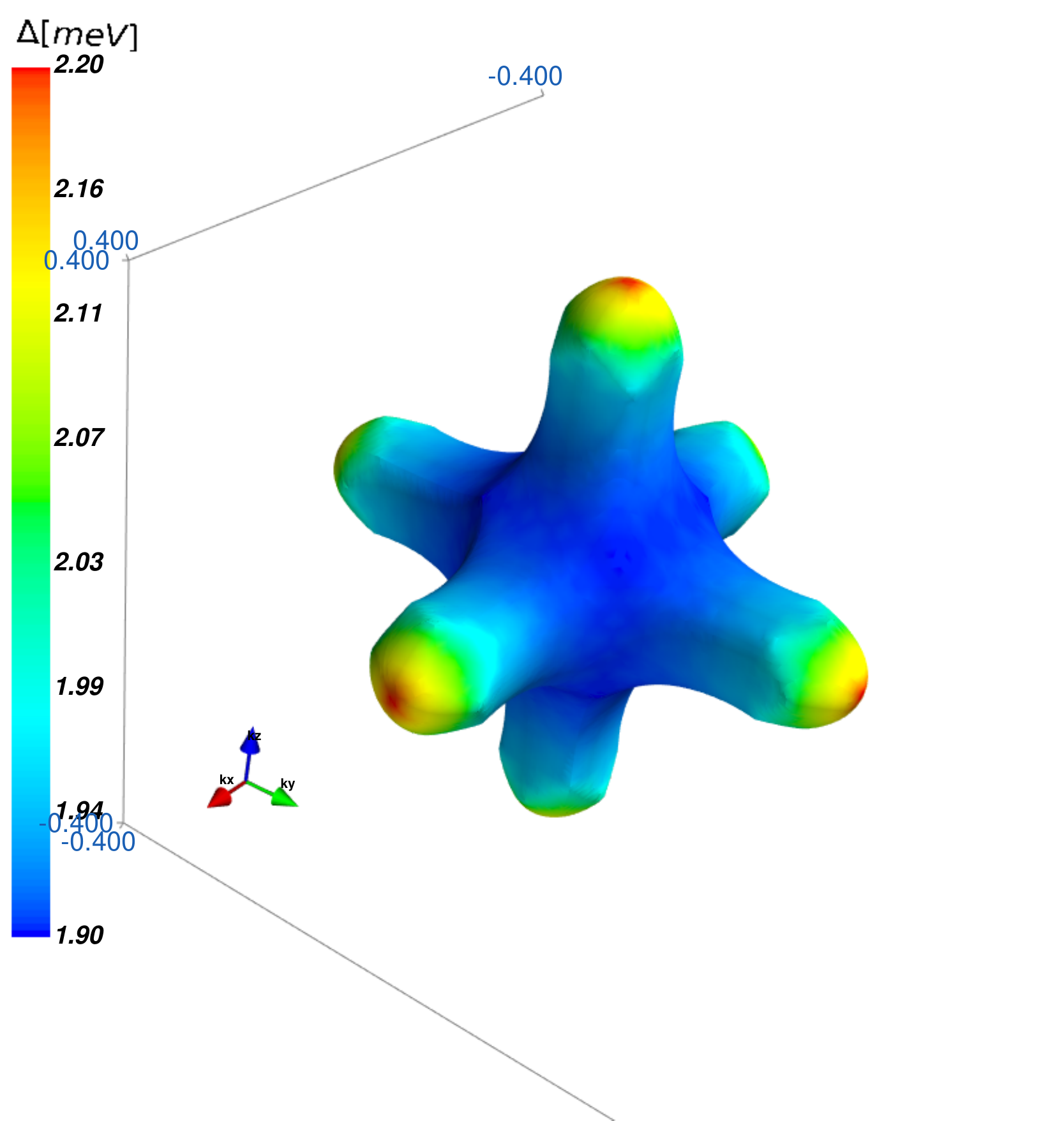}\includegraphics[width=0.5\textwidth]{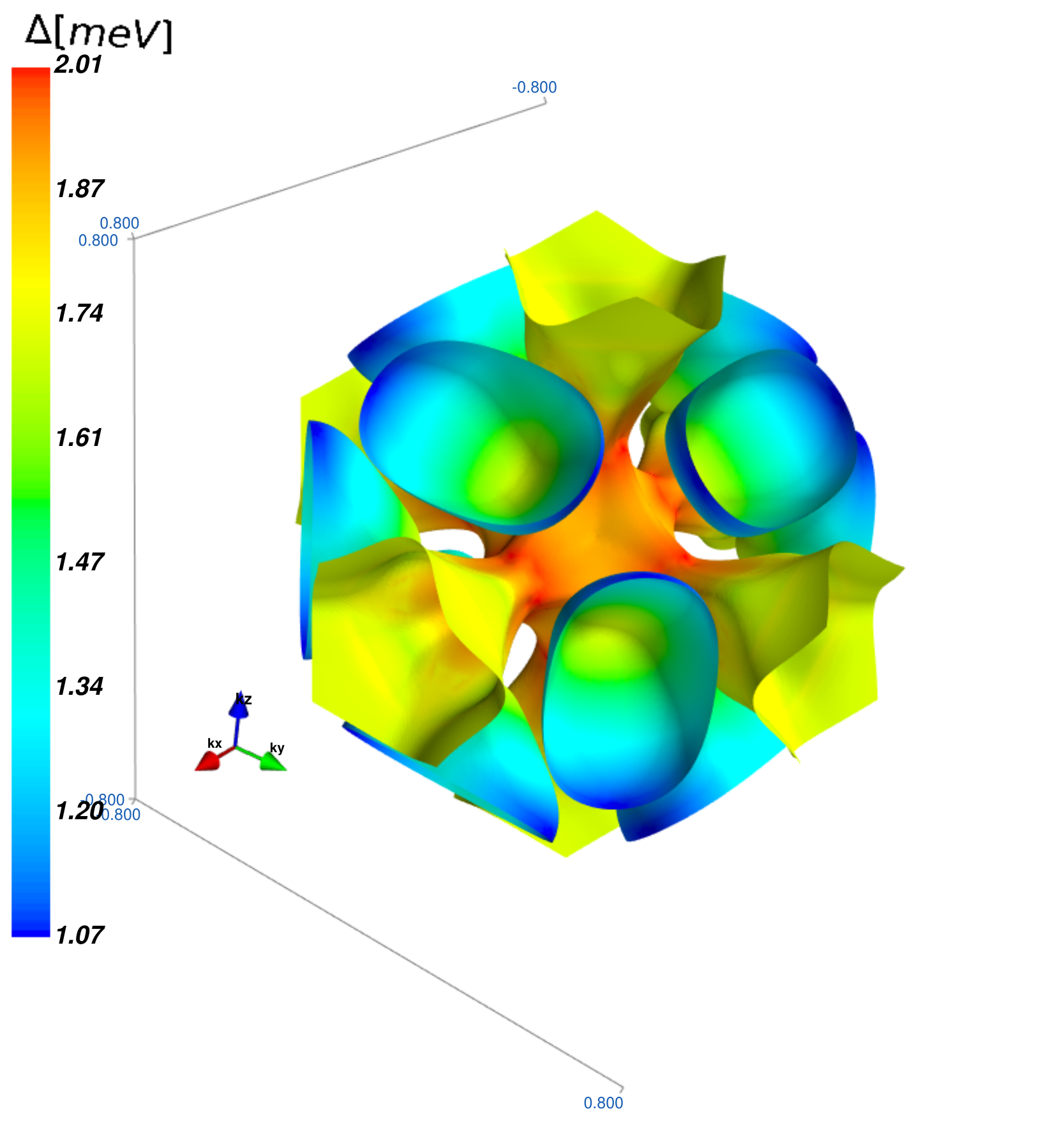}
\caption{The color map indicates the value of $\Delta(\vp)$ at $T=3.0 K$ at each point of the Fermi surface for the two sheets shown in the FBZ. The average gap,\emph{rms} average gap and the dimensionless gap anisotropy parameter are given in Table \ref{table-Delta_anisotropy}.}
\label{fig-Delta_anisotropy_T3.0}
\end{minipage}
\end{figure}
%------------------------------------------------------------------------------

Our results for the magnitude of the gap evaluated at the lowest Matsubara frequency, $\Delta(\vp)\equiv\Delta(\vp;\varepsilon_0=\pi T)$ are shown in Fig.~\ref{fig-Delta_anisotropy_T3.0} for $T=3.0\,\mbox{K}$, $\lambda=1.057$ and $\mu^{\star}=0.218$ for both sheets of the Fermi surface. 
The gap varies from $\Delta_{\text{min}}=1.09\,\mbox{meV}$ to $\Delta_{\text{max}}=2.20\,\mbox{meV}$. However, the maximum and minimum gap values are confined to rather small regions of the Fermi surface. 
Table~\ref{table-Delta_anisotropy} provides a measure of the distribution of gap values on the Fermi surface.
At low temperatures the mean value of the gap averaged over the Fermi surface is dominated by the band 2 with 
$\bar\Delta\equiv\langle\Delta(\vp)\rangle=\int d\vp\,\Delta(\vp)=1.59\,\mbox{meV}$.
The \emph{rms} average gap is slightly higher,
$\Delta_{\text{rms}}\equiv\sqrt{\langle|\Delta(\vp)|^2\rangle}=1.61\,\mbox{meV}$.
This deviation plays an important role in several properties of anisotropic superconductors, including the pair-breaking effect of non-magnetic impurities on the suppression of the superconducting transition temperature discussed in Sec.~\ref{sec-Tc-disorder}.
An important measure of the gap anisotropy is the variance in the gap relative to the average gap normalized to the \emph{rms} average gap, i.e. 
\begin{equation}
\mathcal{A} \equiv 
\frac{\langle|\Delta(\vp)|^2\rangle  - |\langle\Delta(\vp)\rangle|^2}
     {\langle|\Delta(\vp)|^2\rangle}
\end{equation}
Note that this measure of the gap anisotropy varies from $0.028$ at $T=3.0\,\mbox{K}$ to $0.037$ for $T=9.0\,\mbox{K}$, consistent with the expectation based on Eq.~\eqref{eq-Eliashberg-Delta} that the anisotropy is maximun for $T\rightarrow T_c^-$. Thus the variance is
$\sigma_{\text{$\Delta$}}=\sqrt{\cA}\,\Delta_{\text{rms}}\approx 0.19\,\Delta_{\text{rms}}$.
In this limit the anisotropy of $\Delta(\vp)$ reflects the anisotropic eigenfunction, $\cY(\vp)$, for the dominant pairing channel of the linearized gap equation.

%------------------------------------------------------------------------------
\begin{table}[t!]
\begin{center}
\begin{tabular}{|c|c|c|c|c|}
\hline
$T(K)$ 
& 
Band 
& 
$\bar\Delta\equiv\langle\Delta\rangle$ 
& 
$\Delta_{\text{rms}}\equiv\sqrt{\langle|\Delta(\vp)|^2\rangle}$ 
& 
$\mathcal{A}$ 
\\
\hline
3 & 1 & 1.956 & 1.957 & 0.001 \\
\hline
3 & 2 & 1.556 & 1.577 & 0.026 \\
\hline
3 & 1+2 & 1.589 & 1.611 & 0.028 \\
\hline
\hline
9 & 1 & 1.223 & 1.224 & 0.002 \\
\hline
9 & 2 & 0.955 & 0.973 & 0.036 \\
\hline
9 & 1+2 & 0.977 & 0.996 & 0.037 \\
\hline
\end{tabular}
\end{center}
\caption{
The average gap, $\bar\Delta\equiv\langle\Delta\rangle$, and the \emph{rms} average gap, $\Delta_{\text{rms}}\equiv\sqrt{\langle|\Delta(\vp)|^2\rangle}$, on the Fermi surface at $T=3.0 K$ corresponding to Fig.~\ref{fig-Delta_anisotropy_T3.0}.  The data are based on interpolation of the EPW data for the momentum dependence on grids with $k=51^3$ and $q=51^3$.
The lower table is the same for $T=9.0 K$.
}
\label{table-Delta_anisotropy}
\end{table}
%------------------------------------------------------------------------------

\subsection{Anisotropy, Disorder \& Pair-Breaking}\label{sec-Tc-disorder}

Elemental metals such as Al, Nb, Pb, Sn, and Hg are \emph{conventional} superconductors in the sense that the order parameter, $\Delta(\vp)$, reflects the symmetry of the Fermi surface, or equivalently the point group symmetry of the normal metallic phase.
Anisotropy of the gap function, $\Delta(\vp)$, for momenta on the Fermi surface is, in principle, observable in a number of physical properties: anisotropy of the upper critical field, anisotropy of Meissner screening currents with respect to surface and crystal orientation, and more generally the a.c.\  electromagnetic response. 

Elastic scattering by a random potential such as a dilute concentration of impurities embedded in the metal leads to finite lifetimes for the momentum of ballistic quasiparticles and to charge diffusion after several scattering events.
For non-magnetic impurities the transition temperature and excitation gap are unmodified in \emph{isotropic} (``s-wave'') superconductors. This result, widely referred to as ``Anderson's theorem'', arises from the common re-normalization of the spectrum of quasiparticles and Cooper pairs for elastic scattering by the random potential.
However, anisotropy of the pairing interaction, and thus the Cooper pair wave function on the Fermi surface, leads to pair-breaking and violation of the Anderson theorem even for non-magnetic impurities. 
This effect was first studied by Markowitz and Kadanoff~\citep{mar63}, Hohenberg~\citep{hoh64} and Maier~\citep{mai09} for Born scattering by impurities. These authors obtained approximate results for the change in $T_c$ with impurity scattering rate in several limits.
The results we report below provide general results for the pair-breaking suppression of $T_c$ by non-magnetic impurities in anisotropic superconductors, and are not restricted to the Born limit for quasiparticle-impurity scattering, nor to weak anisotropy.

\begin{figure}[t!]
\centering\includegraphics[width=0.75\textwidth]{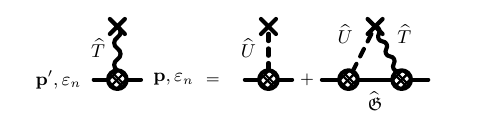}
\caption{Leading-order impurity scattering T-matrix. The internal propagator is the self-consistently determined Nambu propagator.}
\label{fig-impurity-T-matrix}
\end{figure}

The theory of superconducting alloys, as originally formulated by A. Abrikosov and L. Gorkov~\citep{abr59b} and by P. W. Anderson~\citep{and59}, is a generalization of S. Edwards' theory of dilute random impurities in metals~\citep{edw58}. Disorder is treated as a statistical ensemble of random, homogeneously distributed, uncorrelated impurities. 
Thus, to calculate the effects of quasiparticle scattering by a dilute concentration of impurities on the properties of Nb we consider the scattering of quasiparticles and pairs by a static impurity, i.e. the processes represented diagrammatically in Fig.~\ref{fig-impurity-T-matrix}, corresponding to the Bethe-Salpeter equation, 
\ber
\widehat{T}(\vp',\vp;\varepsilon_n) 
=
\widehat{U}(\vp',\vp)
+
N_f\int d\vp''\,\widehat{U}(\vp',\vp'')\widehat{\mathfrak{G}}(\vp'',\varepsilon_n)
\widehat{T}(\vp'',\vp;\varepsilon_n) 
\,.
\label{eq-T-matrix}
\eer
The equation for the Nambu T-matrix describes multiple scattering by a \emph{single} impurity, with the intermediate states defined by the self-consistently determined Nambu matrix propagator. The leading-order electron-impurity self-energy is then given by the T-matrix evaluated in the forward-scattering limit, 
\be\label{eq-Sigma-impurity}
\whSigma_{ei}(\vp;\varepsilon_n)=n_{s}\widehat{T}(\vp,\vp;\varepsilon_n)
\,,
\nonumber
\ee
and the mean impurity density, $n_s$. This mean-field impurity self energy omits contributions from intermediate states involving scattering off more than one impurity. These terms are higher order in the small parameter, $=\hbar/p_f\ell_{\text{ei}}$, where $\ell_{\text{ei}} = v_f\tau_{\text{ei}}$ is the mean-free path for elastic scattering of normal-state quasiparticles by impurities.

%--------------- QP-Impurity T-matrix ----------------------
%------------------------------------------------------------- 

For ``point-like'' impurities only the s-wave scattering channel contributes significantly to the T-matrix, in which case 
\ber
\whT =
\frac{U_0}{1+\pi^2 N_f^2 U_0^2}\,\tone+\frac{N_f U_0^2}{1+\pi^2 N_f^2 U_0^2}\,\langle\whmfG(\vp,\eps_n)\rangle
\,,
\eer
where $U_0$ is the s-wave matrix element of the impurity potential.
The term proportional to the unit matrix does not contribute to static equilibrium properties, and the prefactor of the term proportional to $\langle\whmfG\rangle$ determines the normal-state quasiparticle-impurity scattering rate,
\be
\frac{\hbar}{2\tau} = 
\frac{n_s}{\pi N_f}\,\frac{\pi^2 N_f^2 U_0^2}{1 + \pi^2 N_f^2 U_0^2}
                    = \Gamma\,\sin^2\delta_0 
\,,
\ee
where the second equality is the expression for the scattering rate in terms of the normal-state s-wave scattering phase shift, $\delta_0$, with $\sin\delta_0=\pi N_f U_0/\sqrt{1+\pi^2 N_f^2 U_0^2}$, and $\Gamma = n_s/\pi N_f$ is the scattering rate in the unitarity limit. The total cross section for quasiparticle-impurity scattering is then given by $\sigma=(4\pi\hbar^2/p_f^2)\sin^2\delta_0$. 
We can express the scattering rate as 
$\hbar/2\tau = \nicefrac{4}{3\pi}(\nicefrac{n_s}{n})\,E_f\,\bar\sigma$,
where $n$ is electron density, $E_f$ is the Fermi energy and $\bar\sigma=\sin^2\delta_0$ is the dimensionless cross section normalized to the cross section in the unitarity limit, $\sigma_u=4\pi\hbar^2/p_f^2$. 
Thus, we can express the impurity self energies as,
\ber
\Sigma_{\text{imp}}(\eps_n) 
&=& 
-\frac{\hbar}{2\tau}\,
\left\langle
\frac{i\tilde\eps_n(\vp,\eps_n)}{\sqrt{\tilde\eps_n(\vp,\eps_n)^2 + |\tilde\Delta(\vp,\eps_n)|^2}}
\right\rangle
\,,
\hspace*{5mm}
\\
\Delta_{\text{imp}}(\varepsilon_n) 
&=& 
+\frac{\hbar}{2\tau}\,
\left\langle
\frac{\tilde\Delta(\vp,\eps_n)}{\sqrt{\tilde\eps_n(\vp,\eps_n)^2 + |\tilde\Delta(\vp,\eps_n)|^2}}
\right\rangle
\,.
\eer

In what follows we neglect retardation effects of the electron-phonon coupling. For Nb with $\kb T_c/\omega_c\approx 0.07$~\citep{daa80}, thus retardation effects are relatively small, and that includes the electromagnetic response of Nb at microwave frequencies, $\hbar\omega \ll 2\Delta_0$~\citep{uek22}. At much higher frequencies phonons are observable in the optical absorption spectrum for frequencies just above the gap~\citep{lee89,rai94}. 
Thus, for the purposes of calculating the effects of disorder on the superconducting transition we replace $\lambda(\vp,\vp';\eps_n-\eps_{n'})\rightarrow\lambda(\vp,\vp';0)\Theta(\omega_c-|\eps_n|)\Theta(\omega_c-|\eps_{n'}|)$. 

The anisotropy of the gap function is determined by the momentum dependence of the electron-phonon coupling and angle-resolved quasiparticle density of states at the Fermi energy via the linearized gap equation,
\be\label{eq-LGE}
\Delta(\vp) = \pi T\sum_{\eps_n}^{\omega_c}\int d^2\vp'\,
              \lambda(\vp,\vp')\,\frac{\tilde\Delta(\vp',\eps_n)}{|\tilde\eps_n|}
\,,
\ee
where $\tilde\Delta(\vp,\eps_n)\equiv\Delta(\vp)+\Delta_{\text{imp}}(\varepsilon_n)$ is a linear functional of $\Delta(\vp)$. Note that we have absorbed $-\mu^{\star}$ into $\lambda(\vp,\vp')$. Equation~\eqref{eq-LGE} is an eigenvalue equation for $\Delta(\vp)$ with a spectrum of eigenvalues, $\{T_{c_{\Gamma}}|\Gamma\in\mbox{irrep}\}$, where $T_{c_{\Gamma}}$ is the \emph{instability temperature} for Cooper pair formation with a momentum-space eigenfunctions, $\left\{\cY_{\Gamma,i}(\vp)|i\in d_{\Gamma}\right\}$, belonging to the irreducible representation (irrep) $\Gamma$, of dimension $d_{\Gamma}$, of the crystal point group.
The pairing interaction is invariant under the point group, and thus can be represented as a sum over bilinear products of the eigenfunctions,
\be
\lambda(\vp,\vp') = \sum_{\Gamma}^{\text{irreps}}\,\lambda_{\Gamma}\,
                    \sum_{i=1}^{d_{\Gamma}}\cY_{\Gamma, i}(\vp)\cY^*_{\Gamma, i}(\vp')
\,,
\ee
where $\lambda_{\Gamma}$ is the strength of the pairing interaction in the channel labeled by $\Gamma$. The most attractive interaction determines the highest instability temperature, and thus the superconducting transition temperature, $T_c$. The corresponding eigenfunctions determine the pairing symmetry and gap anisotropy. 

In Eq.~\eqref{eq-LGE} the renormalized Matsubara energy and pairing self energy reduce to
\ber
\tilde\eps_n &=& \eps_n + \sgn(\eps_n)\,\frac{\hbar}{2\tau}
\label{eq-epstilde}
\\
\tilde\Delta(\vp,\eps_n) &=& \Delta(\vp) + \frac{\hbar}{2\tau}\,\frac{\langle\Delta(\vp)\rangle}{|\eps_n|}
\,,
\label{eq-Deltatilde}
\eer
where $1/\tau$ is the normal-state quasiparticle-impurity scattering rate, and $\langle(\ldots)\rangle = \int d\vp(\ldots)$ is the average over the Fermi surface.

We project out the dominant pairing channel and set $\lambda(\vp,\vp')=g\,\cY^*(\vp)\cY(\vp')$, where $g>0$ is the attractive interaction in the dominant pairing channel, and $\cY(\vp)$ is the corresponding eigenfunction.\footnote{For simplicity, and relevance to conventional superconductors such as Nb, we consider only one-dimensional irreps.}
Thus, the order parameter has the form, $\Delta(\vp)=\Delta(T)\,\cY(\vp)$, where $\cY(\vp)$ is normalized, $\langle|\cY(\vp)|^2\rangle = 1$. Using Eqs.~\eqref{eq-epstilde} and \eqref{eq-Deltatilde} we can express Eq.~\eqref{eq-LGE} as an equation for the transition temperature $T_c$ defined in terms of $g$, $\omega_c$ and $\tau$,
\be\label{eq-LGE2}
\hspace*{-3mm}
\frac{1}{g}=\pi T_c\sum_{\eps_n}^{\omega_c}
\frac{1}{|\eps_n|+\frac{\hbar}{2\tau}}
\left(
1 +
|\langle\cY(\vp)\rangle|^2
\times
\frac{\hbar}{2\tau|\eps_n|}
\right)
\,.
\ee
Note that for \emph{conventional} superconductors the superconducting order parameter is in general anisotropic, but retains the full symmetry of the crystal point group, i.e. $\cY(\vp)$ belongs to the \emph{identity} representation with every element $\mathsf C$ of the point group $\mathsf G$ giving $\mathsf{C}\cdot\cY(\vp)=\cY(\vp)$. Thus, for conventional superconductors we have $0 < \langle\cY(\vp)\rangle\le 1$.

It is useful to cast the linearized gap equation as an equation for $T_c$ as a function of $1/\tau$ and the transition temperature, $T_{c_0}$, for pure Nb in the absence of disorder, i.e. $1/\tau =0$; $T_{c_0}$ satisfies, $1/g = K(T_{c_0})$, where
$K(T)\equiv\pi T\sum_{\eps_n}^{\omega_c}\frac{1}{|\eps_n|}\simeq\ln\left(\frac{2e^{\gamma_{\text{E}}}}{\pi}\frac{\omega_c}{T}\right)$.\footnote{$\gamma_{\text{E}}\simeq 0.577216\ldots$ is the Euler-Mascheroni constant.} 
Using $K(T_{c_0})-K(T) = \ln(T/T_{c_0})$ to eliminate $g$ and $\omega_c$ in Eq.~\ref{eq-LGE2} yields
\ber
\ln\left(\frac{T_{c_0}}{T_c}\right) 
&=&
\cA
\,\times
S\left(\frac{\hbar}{2\pi\tau T_c}\right)
\label{eq-Tc_vs_tau}
\\
\mbox{where}\quad  
\cA 
&\equiv&
1 - |\langle\cY(\vp)\rangle|^2
\,,
\eer
is the dimensionless measure of the gap anisotropy at $T_c$. For an isotropic ``s-wave'' superconductor, $\cA=0$, in which case we recover ``Anderson's Theorem'', $T_c=T_{c_0}$~\citep{and59}.
The opposite extreme is the class of \emph{unconventional} superconductors that break the orbital rotation symmetry, and thus belong to one of the non-identity representations of the point group. In this case $\langle\cY(\vp)\rangle \equiv 0$ and $\cA=1$.

%---------------- Tc vs mfp with Gap Anisotropy ----------
\begin{figure}[t!]
\centering\includegraphics[width=0.75\textwidth]{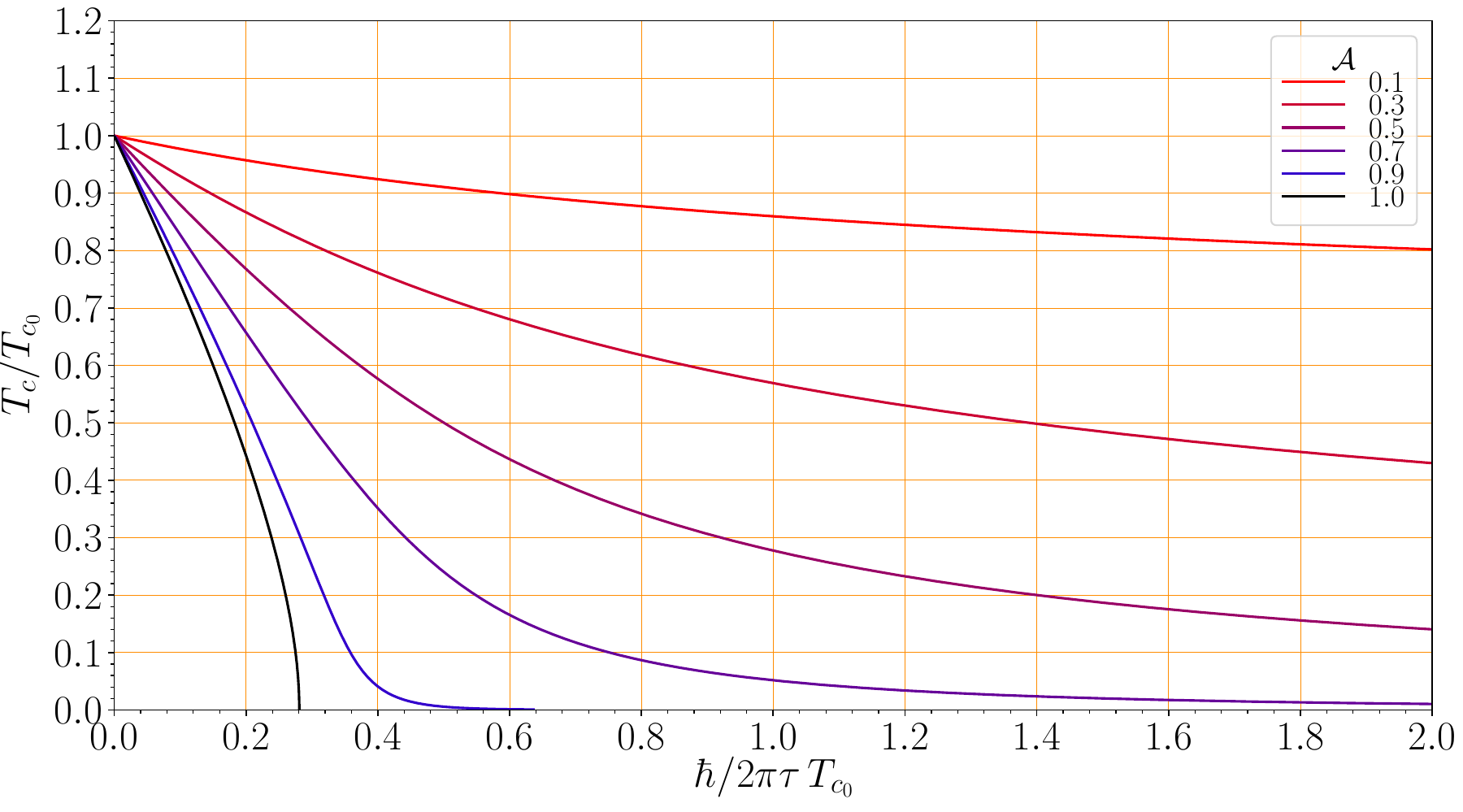}
\caption{Suppression of $T_c$ by disorder over a wide range of possible gap anisotropy values. Note for $\cA=1$ the critical point at which $T_c(\tau_c)=0$ is $\hbar/2\pi\tau_c\,k_B T_{c_0} = 0.281$.
}
\label{fig-Tc_vs_mfp+A}
\end{figure}
%----------------------------------------------------------

Thus, in general anisotropy of the pairing interaction combined with elastic scattering off the disorder potential leads to suppression of the superconducting transition, which is obtained from the solution of Eq.~\ref{eq-Tc_vs_tau} with
\be
S(z) \equiv \sum_{n=0}^{\infty}\frac{\nicefrac{1}{2}z}{(n+\nicefrac{1}{2})(n+\nicefrac{1}{2} + \nicefrac{1}{2}z)}
\,.
\ee
For weak pair-breaking, $\hbar/2\pi\tau T_{c_0}\ll 1$, the suppression of $T_c$ by scattering off the disorder potential becomes,
\be
T_c\simeq T_{c_0}\left(1 - \cA\,\frac{\pi}{8}\,\frac{\hbar}{\tau T_{c_0}}\right)
\,,
\ee
Note that the pair-breaking parameter, $\hbar/2\pi\tau T_{c_0}$, can be expressed as the ratio of the coherence length in the clean limit, $\xi_0=\hbar v_f/2\pi T_{c_0}$, to the transport mean free path, $\ell = v_f\tau$, but what is fundamental is the product of the scattering rate, $1/\tau$, and timescale for Cooper pair formation, $\hbar/2\pi T_{c_0}$. The suppression of $T_c$ by disorder for a wide range of gap anisotropy values is shown in Fig.~\ref{fig-Tc_vs_mfp+A}.
The case $\cA=1$, corresponding to unconventional superconductors with $\langle\cY(\vp)\rangle=0$, is special; $T_c$ vanishes at a disorder critical point given by $\hbar/2\pi\tau_c\,T_{c_0}=\nicefrac{1}{2}e^{-\gamma_{\text{E}}}\simeq 0.281$,
i.e. superconductivity is destroyed for scattering rates, $1/\tau \ge 1/\tau_c$, or equivalently mean free paths, $\ell\le\ell_c=3.56\,\xi_0$.

For anisotropic conventional superconductors with $\langle\cY(\vp)\rangle\ne0$ the transition temperature is suppressed but there is no critical point. Even for relatively weak anisotropy the impact of quasiparticle scattering in the presence of anisotropic pairing can lead to significant suppression of the $T_c$ as is shown in Fig.~\eqref{fig-Tc_vs_mfp+A}. For $\cA=0.1$ $T_c$ is suppressed by $20\%$ for a superconductor with $\hbar/2\pi\tau T_{c_0}=2$, i.e. just into the ``dirty'' regime.

%---------------- Tc vs mfp with Gap Anisotropy ----------
\begin{figure}[t!]
\centering\includegraphics[width=0.875\textwidth]{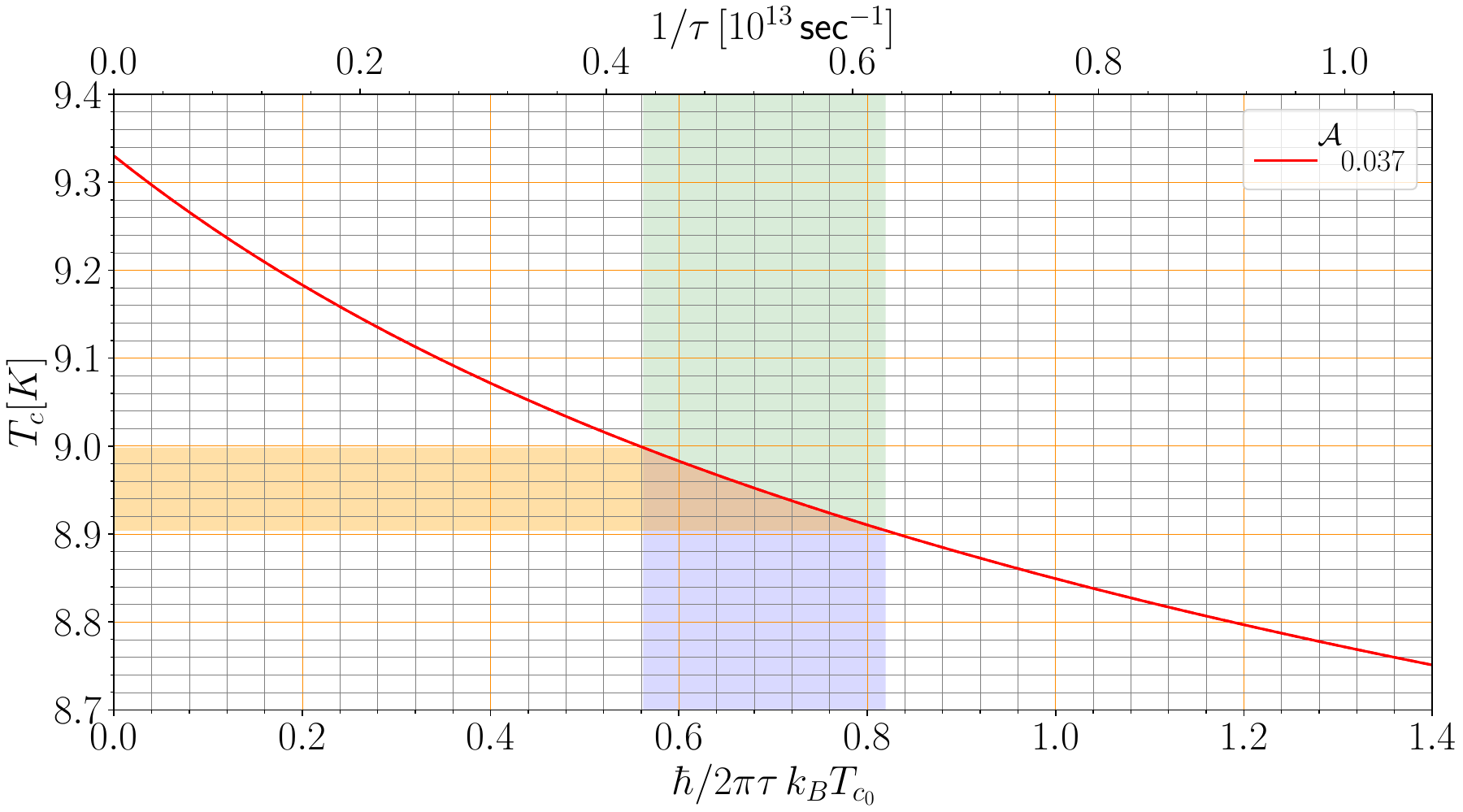}
\caption{Suppression of $T_c$ by disorder from Eq.~\eqref{eq-Tc_vs_tau} for Nb with gap anisotropy parameter $\cA=0.037$ calculated from the eigenfunction, $\cY(\vp)$, of the linearized Eliashberg gap equation in the clean limit, $1/\tau\rightarrow 0$.
}
\label{fig-Tc_vs_mfp+A-Nb}
\end{figure}
%----------------------------------------------------------

\subsection{$T_c$ vs disorder in Nb}

The suppression of $T_c$ for Niobium with the anisotropy ratio calculated from the anisotropic Eliashberg theory using EPW is shown in Fig.~\ref{fig-Tc_vs_mfp+A-Nb}. The combination of scattering by the disorder potential and anisotropy of the electron-phonon coupling can suppress $T_c$ from $T_{c_0}=9.33\,\mbox{K}$ to $T_c\simeq 8.9-9.0\,\mbox{K}$, which corresponds to the transition temperatures of the high-Q Niobium SRF cavities with Nitrogen impurities reported in Refs.~\citep{baf19,baf21,uek22}. Based on the predicted gap anisotropy this level of suppression of $T_c$ implies substantial disorder - below, but approaching, the clean to dirty limit cross-over defined by a quasiparticle-impurity mean scattering time that is approaching the Cooper pair formation time, $\tau\gtrsim \hbar/2\pi\kb T_{c_0}$.

%-----------------------------------------------------------------------------
\section{Conclusion}

A first-principles calculation of superconducting properties of pure Nb single crystals depends 
on accurate determination of the electron-phonon coupling solutions of the anisotropic Eliashberg equations. We used the QE code for BCC Nb to obtain the electron-phonon spectral function that best agrees with available tunneling experiment data. Our result for $\alpha^2F(\omega)$ is good agreement with existing tunneling spectroscopy data except for the spectral weight of the longitudinal phonon peak at $\hbar\omega_{\text{LO}}=23\,\mbox{meV}$.
We obtain an electron-phonon coupling constant of $\lambda=1.057$, renormalized Coulomb interaction, $\mu^{\star}=0.218$ for a transition temperature of $T_c=9.33\,\mbox{K}$. The corresponding strong-coupling gap at $T=0$ is modestly enhanced, $\Delta_0=1.55\,\mbox{meV}$, compared to the weak-coupling BCS value $\Delta_0^{\text{wc}}=1.78\,\kb\,T_c= 1.43\,\mbox{meV}$.
The electron-phonon coupling and superconducting gap for of Nb exhibits substantial anisotropy on Fermi surfaces.
We use these results to predict and analyze the distribution of gap anisotropy and compute the suppression of the superconducting transition temperature using a self-consistent T-matrix theory for quasiparticle-impurity scattering to describe Niobium doped with non-magnetic impurities. 
These results provide a quantitative diagnostic for the level of disorder in high-Q impurity-doped Niobium SRF cavities used for accelerator technology, quantum devices for computing and sensing. 

\section*{Acknowledgments}

We thank Drs. D. Bafia, A. Grassellino and A. Romanenko of Fermilab for discussions, and for sharing their results on high-Q SRF cavities which motivated this study.

\section*{Conflict of Interest Statement}

The authors declare that the research was conducted in the absence of any commercial or financial relationships that could be construed as a potential conflict of interest.

\section*{Author Contributions}

MZ is the primary author responsible for analytical and numerical analysis for this study, and contributed to writing and editing the manuscript.  
HU is contributing author responsible for analysis and editing in this manuscript.
JAS developed the plan for this study, is responsible for the concepts, methodology, and analysis of the role of disorder in this study, project administration, acquisition of funding, authorship and editing of this manuscript. 

\section*{Funding}
The work of MZ and JAS was supported by National Science Foundation Grant PHY-1734332. 
The work of HU was supported by the U.S. Department of Energy, Office of Science, National Quantum Information Science Research Centers, Superconducting Quantum Materials and Systems Center (SQMS) under contract number DE-AC02-07CH11359.

%-----------------------------------------------------------------------------
%\bibliographystyle{Frontiers-Vancouver}
%\bibliography{SRF,CM,Books,Numerics,newref,ref_srf}
%-----------------------------------------------------------------------------

\end{document}